\documentclass[sigconf]{acmart}
\usepackage{enumitem}

\usepackage{booktabs} 
\usepackage{mdframed}
\usepackage{framed}

\usepackage{hyphenat}
\usepackage{fnpct}
\usepackage{algpseudocode}
\usepackage{amsmath,epsfig}
\usepackage[boxed,ruled,vlined,linesnumbered]{algorithm2e}
\usepackage{amsthm}
\usepackage{setspace}
\usepackage{enumitem}
\usepackage{multirow}
\usepackage{hhline}
\usepackage{caption}
\usepackage{subcaption}
\usepackage{graphicx}
\usepackage{wrapfig}
\usepackage{amsmath,epsfig}
\usepackage{amsthm}

\setcopyright{rightsretained}
\usepackage{tikz}
\usepackage{balance}

\usepackage{csquotes}
\usepackage{enumitem}
\usepackage{tablefootnote}
\usepackage{fnpct}
\usepackage{bbm}
\usepackage{lipsum}
\usepackage{setspace}

\makeatletter
\def\thm@space@setup{\thm@preskip=0pt
\thm@postskip=0pt}
\makeatother

\usepackage{array}
\usepackage{makecell}
\renewcommand\theadalign{bc}
\renewcommand\theadfont{\bfseries}
\renewcommand\theadgape{\Gape[4pt]}
\renewcommand\cellgape{\Gape[4pt]}

\newcommand{\B}{\vspace*{-\smallskipamount}}
\newcommand{\BB}{\vspace*{-\medskipamount}}
\newcommand{\BBB}{\vspace*{-\bigskipamount}}

\usepackage{longtable}
\usepackage{mdframed}

\usepackage[hang,flushmargin]{footmisc}

\usepackage{natbib}
\newcommand*\wrapletters[1]{\wr@pletters#1\@nil}
\def\wr@pletters#1#2\@nil{#1\allowbreak\if&#2&\else\wr@pletters#2\@nil\fi}
\usepackage{tikz}
\usetikzlibrary{automata,matrix,shapes,arrows,positioning,chains,calc}
\usetikzlibrary{snakes}
\usetikzlibrary{arrows,scopes}
\usetikzlibrary{positioning,chains,fit,shapes,calc}
\usepackage{caption}
\usepackage{subcaption}
\usepackage{amsmath}
\usepackage{lineno}

\setcopyright{rightsretained}

\newcounter{BalanceAtReference}
\newcounter{ReferenceIndexForBalancing}

\makeatletter

\global\@ACM@balancefalse

\def\@balancelastpageonce{%
  \ifnum\value{ReferenceIndexForBalancing}=\value{BalanceAtReference}
    \newpage
  \else
    \relax
  \fi
  \stepcounter{ReferenceIndexForBalancing}
}
\pretocmd{\bibitem}{\@balancelastpageonce}
  {} 
  {\@latex@error{Patching \bibitem failed}{\@ehd}}

\makeatother

\begin{document}
\fancyhead{}
\title{\textsc{Prism}:
\underline{Pr}ivate Ver\underline{i}fiable \underline{S}et Computation \\ over \underline{M}ulti-Owner Outsourced Databases}\titlenote{\scriptsize This material is based on research sponsored by DARPA under agreement number FA8750-16-2-0021. The U.S. Government is authorized to reproduce and distribute reprints for Governmental purposes notwithstanding any copyright notation thereon. The views and conclusions contained herein are those of the authors and should not be interpreted as necessarily representing the official policies or endorsements, either expressed or implied, of DARPA or the U.S. Government. This work is partially supported by NSF grants 1527536 and 1545071. Yin Li is supported by the National Natural Science Foundation of China (Grant no. 61972090, 61972089, 61601396).}

\author{Yin Li,$^1$ Dhrubajyoti Ghosh,$^2$ Peeyush Gupta,$^2$ Sharad Mehrotra,$^2$ Nisha Panwar,$^3$ \\ Shantanu Sharma$^2$}
\affiliation{\institution{$^1$Dongguan University of Technology, China. $^2$University of California, Irvine, USA. $^3$Augusta University, USA.\\ Email: \texttt{sharad@ics.uci.edu}, \texttt{shantanu.sharma@uci.edu}}}

\setlength{\belowdisplayskip}{0pt}
\setlength{\belowdisplayshortskip}{0pt}
\setlength{\abovedisplayskip}{0pt}
\setlength{\abovedisplayshortskip}{0pt}

\begin{abstract}

This paper proposes \textsc{Prism}, a secret sharing based approach to compute private set operations (\textit{i}.\textit{e}., intersection and union), as well as aggregates over outsourced databases belonging to multiple owners. \textsc{Prism} enables data owners to pre-load the data onto non-colluding servers and exploits the additive and multiplicative properties of secret-shares to compute the above-listed operations in (at most) two rounds of communication between the servers (storing the secret-shares) and the querier, resulting in a very efficient implementation. Also, \textsc{Prism} does not require communication among the servers and supports result verification techniques for each operation to detect malicious adversaries. Experimental results show that \textsc{Prism} scales both in terms of the number of data owners and database sizes, to which prior approaches do not scale. 

\end{abstract}

\copyrightyear{2021}
\acmYear{2021}
\acmConference[SIGMOD '21]{Proceedings of the 2021 International Conference on Management of Data}{June 18--27, 2021}{Virtual Event , China}
\acmBooktitle{Proceedings of the 2021 International Conference on Management of Data (SIGMOD '21), June 18--27, 2021, Virtual Event , China}\acmDOI{10.1145/3448016.3452839}

\acmISBN{978-1-4503-8343-1/21/06}

 \maketitle

\section{Introduction}
\label{sec:Introduction}

With the advent of cloud computing, database-as-a-service (DaS)~\cite{DBLP:conf/sigmod/HacigumusILM02} has gained significant attention. Traditionally, the DaS problem focused on a single database (DB) owner, submitting suitably encrypted data to the cloud over which DB owner (or one of its clients) can execute queries.
{\color{black}A more general use-case is one in which there are multiple datasets, each owned by a different owner. Data owners do not trust each other, but wish to execute queries over common attributes of the dataset. The query execution must not reveal the content of the database belonging to one DB owner to others, except for the leakage that may occur from the answer to the query.} The most common form of such queries is the \emph{private set intersection} (PSI)~\cite{DBLP:conf/eurocrypt/FreedmanNP04}. 
An example use-case of PSI include \emph{syndromic surveillance}, wherein organizations, such as pharmacies and hospitals share information (\textit{e}.\textit{g}., a sudden increase in sales of specific drugs such as analgesics or anti-allergy medicine, telehealth calls, and school absenteeism requests) to enable early detection of community-wide outbreaks of diseases.
PSI is also a building block for performing joins across private databases --- it essentially corresponds to a semi-join operation on the join attribute~\cite{DBLP:conf/sac/Kerschbaum12}.

Private set computations over datasets owned by different DB owners/organizations can, in general, be implemented using secure multiparty computation (SMC)~\cite{DBLP:conf/focs/Yao86,DBLP:conf/stoc/GoldreichMW87,DBLP:journals/iacr/Lindell20}, a well-known cryptographic technique that has been prevalent for more than three decades. SMC allows DB owners to securely execute any function over their datasets without revealing their data to other DB owners. However, SMC can be very slow, often by order of magnitude~\cite{DBLP:conf/ccs/KolesnikovMPRT17}. Consequently, techniques that can more efficiently compute private set operations have been developed; particularly, in the context of PSI and \emph{private set union} (PSU)~\cite{DBLP:conf/crypto/KissnerS05,DBLP:conf/cans/CristofaroGT12}.
PSU refers to privately computing the union of all databases. 
Approaches using homomorphic encryption~\cite{DBLP:conf/ccs/ChenLR17}, polynomial evaluation~\cite{DBLP:conf/eurocrypt/FreedmanNP04},
garbled-circuit techniques~\cite{DBLP:conf/ndss/HuangEK12},
Bloom-filter~\cite{many2012fast}, and
oblivious transfer~\cite{DBLP:conf/uss/Pinkas0Z14,DBLP:conf/crypto/PinkasRTY19} have been proposed to implement private set operations.

Recent work on private set operations has also explored performing aggregation on the result of PSI operations. For instance,~\cite{google} studied the problem of private set intersection sum (PSI Sum), motivated by the internet advertising use-case, where a party maintains information about which customer clicked on specific advertisements during their web session, while another has a list of transactions about items listed in the advertisements that resulted in a purchase by the customers. Both parties might wish to securely learn the total sales that attributed due to customers clicking on advertisements, while neither would like their data to be revealed to the other for reasons including fair/competitive business strategies.

\begin{table*}[!t]
\BB
\scriptsize
\centering
\begin{minipage}{0.32\linewidth}
\centering
\bgroup
\def\arraystretch{.95}
\centering
 \begin{tabular}{|l|l|l|l|l|l|}\hline
      & \texttt{Name} & \texttt{Age} & \texttt{Disease} & \texttt{Cost} \\ \hline\hline
$\tau_1$ & John & 4 & Cancer  & 100  \\ \hline
$\tau_2$ & Adam & 6 & Cancer  & 200  \\ \hline
$\tau_3$ & Mike & 2 & Heart   & 300  \\  \hline
\end{tabular}
 \caption{Hospital 1.}
  \label{tab:table 1}
\egroup
\end{minipage}
\begin{minipage}{0.32\linewidth}
\centering
\bgroup
\def\arraystretch{0.95}
  \centering
\begin{tabular}{|l|l|l|l|l|l|}\hline
      & \texttt{Name} & \texttt{Age} & \texttt{Disease} & \texttt{Cost} \\\hline\hline
$\nu_1$ & John & 8 & Cancer  & 100  \\\hline
$\nu_2$ & Adam & 5  & Fever   & 70   \\\hline
$\nu_3$ & Bob  & 4  & Fever   & 50 \\\hline
\end{tabular}
   \caption{Hospital 2.}
    \label{tab:table 2}
\egroup
\end{minipage}
\begin{minipage}{0.32\linewidth}
\centering
\bgroup
\def\arraystretch{0.95}
  \centering
\begin{tabular}{|l|l|l|l|l|l|} \hline
      & \texttt{Name} & \texttt{Age} & \texttt{Disease} & \texttt{Cost} \\\hline\hline
$\rho_1$ & Carl & 8 & Cancer  & 300  \\\hline
$\rho_2$ & John & 4 & Cancer  & 700   \\\hline
$\rho_3$ & Lisa & 5 & Heart   & 500 \\\hline
\end{tabular}
   \caption{Hospital 3.}
    \label{tab:table 3}
\egroup
\end{minipage}
\BBB\BBB
\\Note: $\tau_i$, $\nu_i$, and $\rho_i$ denote the $i^{\mathit{th}}$ tuples of tables.
\BBB\BB
\end{table*}

Existing approaches on private set computation (including recent work on aggregation) are limited in several ways:

\begin{itemize}[nolistsep,noitemsep,leftmargin=0.01in,topsep=0pt]
\item Work on PSI or PSU has largely focused on the case of two DB owners, with some exceptions that address more than two DB owners scenarios, \textit{e}.\textit{g}.,~\cite{DBLP:conf/ccs/KolesnikovMPRT17,DBLP:conf/scn/InbarOP18,10.1145/3338466.3358927,DBLP:conf/crypto/KissnerS05,DBLP:journals/ieicet/CheonJS12,DBLP:conf/pkc/HazayV17,DBLP:conf/eurocrypt/FreedmanNP04}.  
There are several interesting use-cases, where one may wish to compute PSI over
multiple datasets. For instance, in the syndromic surveillance example listed above, one may wish to compute intersection amongst several independently owned databases. Generalizing existing two-party PSI or PSU approaches to the case of multiple DB owners results in significant overhead~\cite{DBLP:conf/ccs/KolesnikovMPRT17}. For instance,~\cite{DBLP:journals/tdsc/AbadiTMD19}, which is designed for two DB owners, incurs $(nm)^2$ communication cost, when extended to $m>2$ DB owners, where $n$ is the dataset size. 

\item Techniques to privately compute aggregation over set operations have not been studied systematically. In database literature, aggregation functions~\cite{DBLP:conf/osdi/MaddenFHH02} are typically classified as: {\em summary} aggregations (\textit{e}.\textit{g}., count, sum, and average) or {\em exemplary} aggregations (\textit{e}.\textit{g}., minimum, maximum, and median). Existing literature has only considered the problem of
PSI Sum~\cite{google} and cardinality determination, \textit{i}.\textit{e}., the size of the intersection/union~\cite{DBLP:conf/cans/CristofaroGT12,DBLP:conf/acisp/EgertFGJST15}.
Techniques for exemplary aggregations (and even for summary aggregations) that may
compute over multiple attributes have not been explored.

\item Many of the existing solutions do not deal with a large amount of data, due to either inefficient cryptographic techniques or multiple communication rounds amongst DB owners. For instance, recent work~\cite{10.1145/3338466.3358927,DBLP:conf/ccs/KolesnikovMPRT17,DBLP:conf/ccs/LeRG19} dealt with data
that is limited to sets of size less than or equal to $\approx$1M in size. 
\end{itemize}

This paper introduces \textsc{Prism} --- a novel approach for computing collaboratively over multiple databases. \textsc{Prism} is designed for both PSI and PSU, and supports both summary, as well as, exemplar aggregations. Unlike existing SMC techniques (wherein DB owners compute operations privately through a sequence of communication rounds), in \textsc{Prism}, DB owners outsource their data in secret-shared form to multiple \textbf{\emph{non-communicating public servers}}. As will become clear, \textsc{Prism} exploits the homomorphic nature of secret-shares to enable servers to compute private set operations independently (to a large degree). These results are then returned to DB owners to compute the final results. In \textsc{Prism}, any operator requires at most two communication rounds between DB owners and servers, where the first round finds tuples that are in the intersection or union of the set, and the second round computes the aggregation function over the objects in the intersection/union.




By using public servers for computation over secret-shared data, \textsc{Prism} achieves the identical security guarantees as existing SMC systems (\textit{e}.\textit{g}., Sharemind~\cite{Sharemind}, Jana~\cite{DBLP:journals/iacr/ArcherBLKNPSW18}, and Conclave~\cite{DBLP:conf/eurosys/VolgushevSGVLB19}). The key
advantage of \textsc{Prism} is that by outsourcing data in secret shared form and exploiting homomorphic
properties, \textsc{Prism} does not require communication among server before/during/after the computation, which allows \textsc{Prism} to perform efficiently even for large data sizes and for a large number of DB owners (as we will show in experiment section). Since \textsc{Prism} uses the public servers, which may act maliciously, \textsc{Prism} supports oblivious result verification methods.

\noindent
\textbf{Advantages of \textsc{Prism}.}
In summary, \textsc{Prism} offers the following benefits:
(\textit{i}) \emph{Information-theoretical security}: It achieves information-theoretical security at the servers and prevents them to learn anything from input/output/access-patterns/output-size.
(\textit{ii}) \emph{No communication among servers}: It does not require any communication among servers, unlike SMC-based solutions.
(\textit{iii}) \emph{No trusted entity}: It does not require any trusted entity that performs the computation on the cleartext data, unlike the recent SMC system Conclave~\cite{DBLP:conf/eurosys/VolgushevSGVLB19}.
(\textit{iv}) \emph{Several DB owners and large-sized dataset}: It deals with several DB owners having a large-size dataset.


\noindent
\textbf{Full version~\cite{full_version}.} provides result verification methods for different aggregation approaches, correctness, and information leakage discussions. 

\section{Private Set Operations}
\label{sec:Private Set Operations}
We, first, define the set of operations supported by \textsc{Prism}. Let $\mathit{DB}_1,\ldots, \mathit{DB}_m$ ($m>2$) be independent DBs owned by $m$ DB owners $\mathcal{DB}_1,\ldots,\mathcal{DB}_m$.
{\color{black} We assume, each DB owner is (partially) aware of the schema of data stored at other DB owners. Particularly, DB owners have knowledge of attribute(s) of the data stored at other DB owners on which the set-based operations (intersection/union) can be performed. Also, DB owners know about the attributes on which aggregation functions be supported. This assumption is needed to ensure that PSI/PSU and aggregation queries are well defined. However, the schema of data at different databases may be different.}

Now, we define the private set operations supported by \textsc{Prism} formally and their corresponding privacy requirements (corresponding SQL statements are shown in Table~\ref{tab:SQL syntax of operations supported}). To do so, (and in the rest of the paper), we use the example tables shown in Tables~\ref{tab:table 1},~\ref{tab:table 2}, and~\ref{tab:table 3} that are owned by three different DB owners (in our case, hospitals).

\begin{enumerate}[nolistsep,noitemsep,leftmargin=0.01in, topsep=0pt]
\item \noindent\textbf{\textit{Private set intersection (PSI)} (\S\ref{sec:Common Item Finding}).} PSI finds the common values among $m$ DB owners for a specific attribute $A_c$, \textit{i}.\textit{e}., $\mathit{DB}_1.A_c\cap \ldots \cap \mathit{DB}_m.A_c$. For example, PSI over disease column of Tables~\ref{tab:table 1},~\ref{tab:table 2}, and~\ref{tab:table 3} returns $\{$Cancer$\}$ as a common disease treated by all hospitals. Note that \emph{a hospital computing PSI on disease should not gain any information about other possible disease values (except for the result of the PSI) associated with other hospitals}.

\item \textbf{\textit{Private set union (PSU)} (\S\ref{sec:Generalized Union Query}).} PSU finds the union of values among $m$ DB owners for a specific attribute $A_c$, \textit{i}.\textit{e}., 
$\mathit{DB}_1.A_c\cup \ldots \cup \mathit{DB}_m.A_c$. E.g., PSU over disease column returns $\{$Cancer, Fever, Heart$\}$ as diseases treated by all hospitals. 
\emph{A hospital computing PSU over other hospitals must not gain information about the specific diseases treated by others, or how many hospitals treat which disease}.

\item \noindent \textbf{\textit{Aggregation over private set operators (\S\ref{sec:Extending PSI}.)}}
Aggregation $_{A_c}\mathcal{G}_{\theta}(A_x)$ 
computes an aggregation function $\theta$ on attribute $A_x$ ($A_c$ $\neq$ $A_x$) for the groups corresponding to the output of set-based operations (PSI/PSU) on attribute $A_c$. E.g., the aggregation function $\mathit{sum}$ on cost attribute corresponding to PSI over disease attribute (\textit{i}.\textit{e}., $_{\textnormal{disease}}\mathcal{G}_{sum}(\textnormal{cost})$) returns a tuple $\{$Cancer,1400$\}$. The same aggregation function over PSU will return $\{\langle$Cancer,1400$\rangle,
\langle$Fever,120 $\rangle$, $\langle$ Heart,800$\rangle\}$. Likewise, the output of aggregation  $_{\textnormal{disease}}\mathcal{G}_{max}(\textnormal{age})$ over PSI would return $\{$Cancer,8$\}$, while the same over PSU would return $\{\langle$Cancer,8$\rangle,\langle$Fever,5$\rangle, \langle$Heart,5$\rangle\}$. Note that the count operation does not require specifying an aggregation attribute $A_x$ and can be computed over the attribute(s) associated with PSI/PSU. E.g., count over PSI (PSU) on disease column will return 1 (3), respectively. 
From the perspective of privacy requirement, in the case of PSI on disease column, a hospital executing an aggregation query (maximum of age or sum of cost) should only gain information about the answer, \textit{i}.\textit{e}., \emph{elements in the PSI and the corresponding aggregate value}. It should not gain information about other diseases that are not in the intersection. Likewise, for PSU, the hospital will gain information about \emph{all elements in the union and their corresponding aggregate values, but will not gain any specific information about which database contains which disease values, or the number of databases with a specific disease}.
\end{enumerate}

\begin{table*}[!t]
\BB
\scriptsize
\centering
\begin{tabular}{|l|l|}\hline
  PSI & \texttt{SELECT $A_c$ FROM $db_1$
INTERSECT
$\ldots$
INTERSECT
SELECT $A_c$ FROM $\mathit{db}_m$} \\\hline

PSU & {
\texttt{SELECT $A_c$ FROM $\mathit{db}_1$
UNION
$\ldots$
UNION
SELECT $A_c$ FROM $\mathit{db}_m$}} \\\hline

PSI count &
\texttt{SELECT COUNT($A_c$) FROM $db_1$
INTERSECT
$\ldots$
INTERSECT
SELECT $A_c$ FROM $\mathit{db}_m$} \\\hline

\makecell[l]{PSI $\mathit{\theta}$\\
$ \mathit{\theta \in}$ (AVG, SUM, MAX, MIN, Median)} & \makecell{
\texttt{CREATE VIEW $\mathit{CommonA_c}$ as
SELECT $A_c$ FROM $\mathit{db}_1$
INTERSECT
$\ldots$
INTERSECT
SELECT $A_c$ FROM $\mathit{db}_m$} \\
\texttt{
SELECT $A_c$, $\mathit\theta$($A_x$) FROM
(SELECT $A_x$, $A_c$ FROM $\mathit{db}_1$, $\mathit{CommonA_c}$ WHERE $\mathit{db_1.A_c}$ = $\mathit{CommonA_c.A_c}$
UNION ALL
$\ldots$
UNION ALL}\\
\texttt{SELECT $A_x$, $A_c$ FROM $\mathit{db}_m$, $\mathit{CommonA_c}$ WHERE
$\mathit{db_m.A_c}$ = $\mathit{CommonA_c.A_c}$) as inner\_relation
Group By $A_c$}}
 \\\hline

\end{tabular}
\caption{SQL syntax of operations supported by \textsc{Prism}.}
\label{tab:SQL syntax of operations supported}
\BBB\BBB\B\BBB
\end{table*}

\section{Preliminary}
\label{sec:preliminary}
This section presents the cryptographic concepts that serve as building blocks for \textsc{Prism}, an overview of \textsc{Prism}, and security properties.

\subsection{Building Blocks}
\label{subsec:Building Blocks}
\textsc{Prism} is based on additive secret-sharing (SS), Shamir's secret-sharing (SSS), cyclic group, and pseudorandom number generator. We provide an overview of these techniques, below.

\smallskip
\noindent\textbf{Additive Secret-Sharing (SS).} Additive SS is the simplest type of the SS. Let $\delta$ be a prime number. Let $\mathbb{G}_{\delta}$ be an Abelian group under modulo addition ${\delta}$ operation.
All additive shares are defined over $\mathbb{G}_{\delta}$.
In particular, the DB owner creates $c$ shares $A(s)^1, A(s)^2, \ldots, A(s)^c$ over $\mathbb{G}_{\delta}$ of a secret, say $s$, such that $s=A(s)^1+ A(s)^2+ \ldots+ A(s)^c$. The DB owner sends share $A(s)^i$ to the $i^{\mathit{th}}$ server (belonging to a set of $c$ non-communicating servers). These servers cannot know the secret $s$ until they collect all $c$ shares. To reconstruct $s$, the DB owner collects all the shares and adds them. Additive SS allows \emph{additive homomorphism}.
Thus, servers holding shares of different secrets can locally compute the sum of those shares. Let $A(x)^i$ and $A(y)^i$ be additive shares of two secrets $x$ and $y$, respectively, at a server $i$, then the server $i$ can compute $A(x)^i+A(y)^i$ that enable DB owner to know the result of $x+y$. The precondition of \emph{additive homomorphism is that the sum of shares should be less than $\delta$}.

\noindent\textit{\underline{Example.}} Let $\mathbb{G}_5=\{0, 1, 2, 3, 4\}$ be an
Abelian group under the addition modulo $5$. Let $4$ be a secret. A DB owner may create two shares: $3$ and $1$ (since $4=(3+1)\bmod 5$). 

\smallskip
\noindent\textbf{Shamir's Secret-Sharing (SSS)~\cite{DBLP:journals/cacm/Shamir79}.} Let $s$ be a secret. A DB owner randomly selects a polynomial of degree $c^{\prime}$ with $c^{\prime}$ random coefficients, \textit{i}.\textit{e}., $f(x)= a_0+a_1x+a_2x^2+\cdots+a_{c^{\prime}}x^{c^{\prime}}$, where $f(x)\in \mathbb{F}_p[x]$, $p$ is a prime number, $\mathbb{F}_p$ is a finite field of order $p$, $a_0=s$, and $a_i\in \mathbb{N}$ ($1\leq i\leq c^{\prime}$). The DB owner distributes $s$ into $c$ shares by computing $f(x)$ ($x=1,\ldots, c$) and sends an $i^{\mathit{th}}$ share to an $i^{\mathit{th}}$ server (belonging to a set of $c$ non-colluding servers). The secret can be reconstructed using any $c^{\prime}+1$ shares using Lagrange interpolation~\cite{corless2013graduate}. 
SSS allows \textit{additive homomorphism}, \textit{i}.\textit{e}., if $S(x)^i$ and $S(y)^i$ are SSS of two secrets $x$ and $y$, respectively, at a server $i$, then the server $i$ can compute $S(x)^i+S(y)^i$, which will result in $x+y$ at DB owner.

\smallskip
\noindent\textbf{Cyclic group under modulo multiplication.} Let $\eta$ be a prime number. A group $\mathbb{G}$ is called a cyclic group, if there exists an element $g \in \mathbb{G}$, such that all $x \in \mathbb{G}$ can be derived as $x = (g^i)$ (where $i$ in an integer number $\mathbb{Z}$) under modulo multiplicative $\eta$ operation. The element $g$ is called a generator of the cyclic group. The number of elements in $\mathbb{G}$ is called the \emph{order} of $\mathbb{G}$. Based on each element $x$ of a cyclic group, we can form a cyclic subgroup by executing $x^i\bmod \eta$. 

\noindent\textit{\underline{Example.}}
$g=2$ is a generator of a cyclic group under multiplication modulo $\eta=11$ for the group: $\{1,2,3,4,5,6,7,8,9,10\}$. Note that the group elements are derived by $2^i\bmod 11$. By taking the element 5 of this cyclic group, we form the following cyclic subgroup $\{1,3,4,5,9\}$, under multiplication modulo $\eta=11$, by $5^i\bmod 11$.

\smallskip
\noindent\textbf{Permutation function $\mathcal{PF}$.} Let $A$ be a set. A permutation function $\mathcal{PF}$ is a bijective function that maps a permutation of $A$ to another permutation of $A$, \textit{i}.\textit{e}.,  $\mathcal{PF}: A\rightarrow A$.

%

\smallskip
\noindent\textbf{Pseudorandom number generator $\mathcal{PRG}$:} is a deterministic and efficient algorithm that generates a pseudorandom number sequence based on an input seed~\cite{DBLP:journals/siamcomp/BlumM84,DBLP:journals/jacm/GoldreichGM86}.


\subsection{Entities and Trust Assumption}
\label{subsec:Entities}

\textsc{Prism} assumes the following four entities:

\begin{enumerate}[noitemsep,nolistsep,leftmargin=0.01in]
\item
The $m$ \textbf{database (DB) owners} (or users), who wish to execute computation on their joint datasets. We assume that each DB owner is trusted and 
does not act maliciously. 

\smallskip
\item
{\color{black} A set of \textbf{\emph{$c \geq 2$ servers}} that store the secret-shared data outsourced by DB owners and execute the requested computation from authenticated DB owners. Data transmission between a DB owner and a server takes place in encrypted form or using anonymous routing~\cite{DBLP:journals/cacm/GoldschlagRS99} to prevent the locations of all servers from an adversary. 

We assume that servers do not maliciously communicate (\textit{i}.\textit{e}., non-communicating servers) with each other in violation of \textsc{Prism} protocols. Unlike other MPC mechanisms~\cite{Sharemind}, (as will be clear soon), \textsc{Prism} protocols do not require the servers to communicate before/during/after the execution of the query. The security of secret-sharing techniques requires that out of the $c$ servers, no more than $c^{\prime} < c$ communicate maliciously or collude with each other, where $c^{\prime}$ is a minority of servers (\textit{i}.\textit{e}., less than half of $c$). Thus, we assume that a majority of servers do not collude and communicate with each other, and hence, a legal secret value cannot be generated/inserted/updated/deleted at the majority of the servers.

Also, note that the collusion of servers in violation of the protocol is a general requirement for secret-sharing based protocols, and a similar assumption is made by many prior  work~\cite{DBLP:journals/cacm/Shamir79,Sharemind,DBLP:conf/nsdi/WangYGVZ17,DBLP:journals/iacr/ChidaHIKKP19}. This assumption is based on  factors such as economic incentivization (violation is against their economic interest), law (illegal to collude), and  jurisdictional boundaries. Such servers can be selected on different clouds, which make the assumption more realistic.

For the purpose of simplicity, we assume, none of the servers colludes with each other, \textit{i}.\textit{e}., they do not communicate directly. Thus, to reconstruct the original secret value from the shares, \emph{two additive shares} suffice. In the case of PSI sum (as  in~\S\ref{subsec:psi_Sum Query}), we need to multiply two shares, each of degree one, and that increases the degree of the polynomial to two. To reconstruct the secret value of degree two, we need at least three multiplicative shares.

While we assume that servers do not collude, we consider two types of adversarial models for servers in the context of the computation that they perform:
    (\textit{i}) \textbf{\emph{Honest-but-curious}} (HBC) servers: correctly compute the assigned task without tampering with data or hiding answers. It may exploit side information (\textit{e}.\textit{g}., the internal state of the server, query execution, background knowledge, and output size) to gain  information about stored data, computation, or results. HBC adversarial model is considered widely in many cryptographic algorithms~\cite{DBLP:conf/stoc/CanettiFGN96,DBLP:conf/sigmod/HacigumusILM02,DBLP:conf/icdcs/WangCLRL10}.
 %
 (\textit{ii}) \textbf{\emph{Malicious}} adversarial servers: can delete/insert tuples from the relation, and hence, is a stronger adversarial model than HBC.
}

\smallskip
\item
An \textbf{\emph{initiator} or \emph{oracle}}, who knows $m$ DB owners and servers. {\color{black}Before data outsourcing by DB owners, the initiator informs the identity of servers to DB owners and vice versa.} Also, the initiator informs the desired parameters (\textit{e}.\textit{g}., a hash function, parameters related to Abelian and cyclic groups, $\mathcal{PF}$, and $\mathcal{RRG}$) to servers and DB owners. {\color{black} The initiator is an entity trusted by all other entities and plays a role similar to the trusted certificate authority in the public-key infrastructure. The initiator never knows the data/results, since it does not store any data, or data/results are not provided to servers via the initiator.} The role of the initiator has also been considered in existing PSI work~\cite{DBLP:journals/tcc/QiuLSLW18,DBLP:conf/ic2e/ZhengX15}.

\smallskip
\item
An \textbf{announcer $\mathcal{S}_a$} who participates only in maximum,  minimum, and median queries to announce the results. $\mathcal{S}_a$ communicates (not maliciously) with servers and initiator (and not with DB owners).
\end{enumerate}


\subsection{ \textsc{Prism} Overview}
\label{subsec:Prism Overview}

\begin{wrapfigure}{r}{0.18\textwidth}
\BBB\BB
\centering
\includegraphics[scale=0.35]{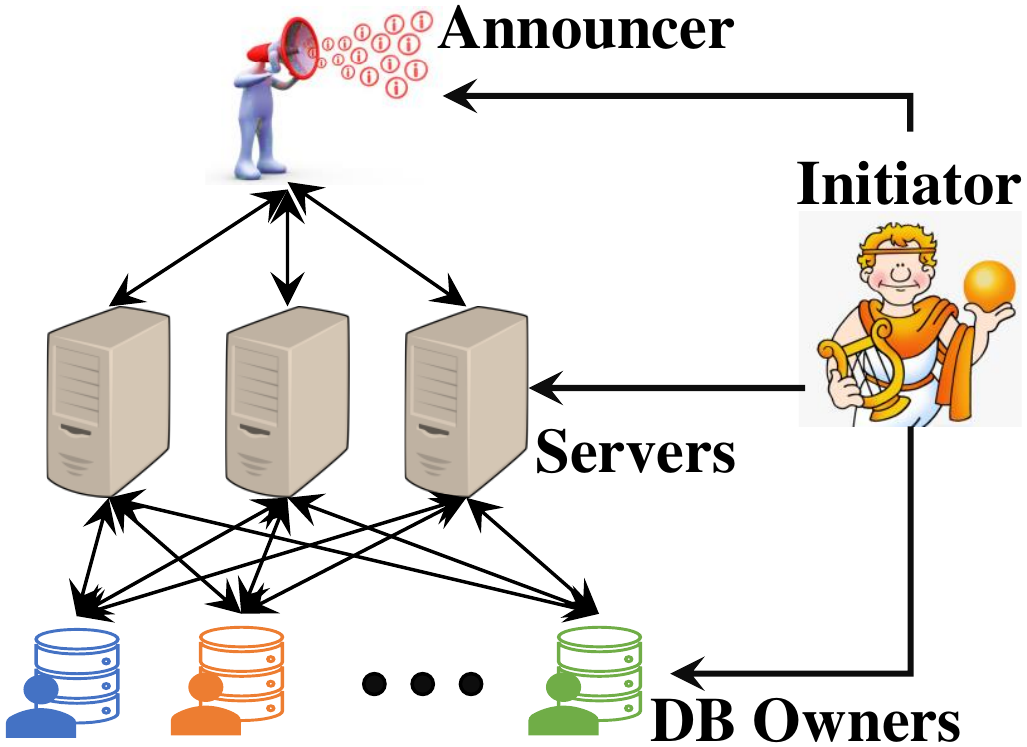}
\caption{\textsc{Prism} model.}
\label{fig:model}
\BBB\B
\end{wrapfigure}

Let us first understand the working of \textsc{Prism} at the high-level. \textsc{Prism} contains four phases {\color{black}(see Figure~\ref{fig:model})}, as follows:

\noindent
\textbf{\textsc{Phase} 0: \emph{Initialization.}} The initiator sends  desired parameters (see details in~\S\ref{sec:Assumptions Related to Parameters}) related to additive SS, SSS, cyclic group, $\mathcal{PF}$, and $\mathcal{PRG}$ to all entities and informs them about the identity of others from/to whom they will receive/send the data.

\setlength{\columnsep}{6pt}

\noindent
\textbf{\textsc{Phase} 1: \emph{Data Outsourcing by DB owners.}} DB owners create additive SS or SSS of their data, by following methods given in \S\ref{sec:Common Item Finding} for PSI and PSU, \S\ref{subsec:psi_Sum Query} for PSI/PSU-sum, and \S\ref{subsec:psi_maximum} for PSI/PSU-max/min. Then, they outsource their secret-shared data to non-communicating servers. Note that in our explanations, we will write the data outsourcing method along with query execution.

\noindent
\textbf{\textsc{Phase} 2: \emph{Query Generation by the DB owner.}} A DB owner who wishes to execute SMC over datasets of different DB owners, sends the query to the servers. For generating secret-shared queries for PSI, PSU, count, sum, maximum, and for their verification, the DB owner follows the method given in \S\ref{sec:Common Item Finding},~\S\ref{sec:Extending PSI}.

\noindent
\textbf{\textsc{Phase} 3: \emph{Query Processing.}} Servers process an input query and respective verification method in an oblivious manner. Neither the query nor the results satisfying the query/verification are revealed to the server. Finally,  servers transfer their outputs to DB owners.

\noindent
\textbf{\textsc{Phase} 4: \emph{Final processing at the DB owners.}} The DB owner either adds the additive shares or performs Lagrange interpolation on SSS to obtain the answer to the query.

\subsection{Security Property}
\label{subsec:Security Property}%

As mentioned in the adversarial setting in \S\ref{subsec:Entities}, an adversarial server wishes to learn the (entire/partial) input and output data, while a DB owner may wish to know the data of other DB owners. Thus, a secure algorithm must prevent an adversary to learn the data (\textit{i}) from the ciphertext representation of the data, (\textit{ii}) from query execution due to access-patterns (\textit{i}.\textit{e}., {\color{black} the adversary can learn the physical locations of tuples that are accessed to answer the query}), and (\textit{iii}) from the size of the output (\textit{i}.\textit{e}., the adversary can learn the number of tuples satisfy the query). The attacks on a dataset based on access-patterns and output-size are discussed in~\cite{DBLP:conf/ndss/IslamKK12,DBLP:conf/ccs/CashGPR15}. In order to prevent these attacks, our security properties are identical to the standard security definition as in~\cite{DBLP:conf/stoc/CanettiFGN96,DBLP:journals/joc/Canetti00,DBLP:conf/tcc/FreedmanIPR05}.
An algorithm is \emph{privacy-preserving} if it maintains DB owners' privacy, data/computation privacy from the servers, and performs identical operations regardless of the inputs.

\smallskip
\noindent\textbf{Privacy from servers} requires that datasets of DB owners must be hidden from servers, before/during/after any computation.
In PSI/ PSU, servers must not know whether a value is common or not, the number of DB owners having a particular value in the result set. In the case of aggregation operations, the output of aggregation over an attribute $A_x$ corresponding to the attributes $A_c$ involved in PSI or PSU should not be revealed to servers. Also, in the case of max/median/min query, servers must not know the max/median/min value and the identity of the DB owner who possesses such values.
Further, the protocol must ensure that the server's behavior in reading/sending the data must be identical for a particular type of query (\textit{e}.\textit{g}., PSI or PSU), thereby the server should not learn anything from query execution (\textit{i}.\textit{e}., hiding access-patterns and output-sizes).

\smallskip
\noindent\textbf{DB owner privacy} requires that the DB owners must not learn anything other than their datasets and the final output of the computation. 
For example, in PSI/PSU queries, DB owners must only learn the intersection/union set, and they must not learn the number of DB owners that does not
contain a particular value in their datasets. Similarly, in the case of aggregation operations, DB owners must only learn the output of aggregation operation, not the individual values on which aggregation was performed.

\smallskip
\noindent\textbf{Properties of verification.} 
A verification method must be oblivious and find misbehavior of servers in computing a query. We follow the verification properties from~\cite{murat08} that the verification method cannot be refuted by the majority of the  servers and should not leak any additional information.

{\color{black}

\section{Assumptions \& Parameters}
\label{sec:Assumptions Related to Parameters}
Different entities in \textsc{Prism} protocols are aware of the following parameters to execute the desired task:

\smallskip
\noindent\textbf{Parameters known to the initiator.}
{\color{black} The initiator knows all parameters used in \textsc{Prism} and distributes them to different entities (only once) as they join in PRISM protocols. Note that the initiator can select these parameters (such as $\eta$, $\delta$) to be large to support increasing DB owners over time without updating parameters. Thus, when new DB owners join, the initiator simply needs to inform DB owners/servers about the increase in the number of DB owners in the system, but does not need to change \emph{all} parameters.}

Additionally, the initiator does the following:
%
(\textit{i}) Selects a polynomial ($\mathcal{F}(x)=a_{m+1}x^{m+1}+a_{m}x^{m}+\ldots+a_1x+a_0$, where $a_i>0$) of degree more than $m$, where $m$ is the number of DB owners, and sends the polynomial to all DB owners. This polynomial will be used during the maximum computation. Importantly, this polynomial $\mathcal{F}(x)$ generates values at different DB owners in an order-preserving manner, as will be clear in~\S\ref{subsec:psi_maximum}, and the degree of the polynomial must be more than $m$ to prevent an entity, who has $m$ different values generated using this polynomial, to reconstruct the secret value (a condition similar to SSS); and beyond $m+1$, the degree of the polynomial does not impact the security, in this case.
%
(\textit{ii}) Generates a permutation function $\mathcal{PF}_i$, and produces four different permutation functions that satisfy Equation~\ref{eq:permutation}:
\begin{equation}\label{eq:permutation}
\mathcal{PF}_{s1} \odot \mathcal{PF}_{db1}=\mathcal{PF}_{s2} \odot \mathcal{PF}_{db2}=\mathcal{PF}_i
\end{equation}
Symbol $\odot$ shows composition of permutations, and these functions can be selected over a permutation group. The initiator provides $\mathcal{PF}_{s1}$, $\mathcal{PF}_{s2}$ to all servers and $\mathcal{PF}_{db1}$, $\mathcal{PF}_{db2}$ to all DB owners.

\medskip
\noindent\textbf{Parameters known to announcer.} Announcer $\mathcal{S}_a$ knows  $\delta$, a prime number used in modulo addition for an Abelian group (\S\ref{subsec:Building Blocks}). 

\medskip
\noindent\textbf{Parameters known to DB owners.} All DB owners know the following parameters:
(\textit{i}) $m$, \textit{i}.\textit{e}., the number of DB owners.
%
(\textit{ii}) $\delta>m$, 
%
(\textit{iii}) $\eta$, where $\eta$ is a prime number used to define modular multiplication for a cyclic group (\S\ref{subsec:Building Blocks}). Note that DB owners do not know the generator $g$ of the cyclic group.
%
(\textit{iv}) A common hash function.
%
(\textit{v}) The domain of the attribute $A_c$ on which they want to execute PSI/PSU. Note that knowing the domain of the attribute $A_c$ does not reveal that which of the DB owner has a value of the domain. (Such an assumption is also considered in prior work~\cite{DBLP:conf/ndss/HuangEK12}.)
%
 %
(\textit{vi}) Two permutation functions $\mathcal{PF}_{\mathit{db1}}$ and $\mathcal{PF}_{\mathit{db2}}$. 
%
(\textit{vii}) The polynomial $\mathcal{F}(x)$ given by the initiator.
(\textit{viii}) A permutation function $\mathcal{PF}$, and the same permutation function will also known to servers.

PSI, PSU, sum, average, count algorithms are based on the assumptions 1-5. PSI verification, sum verification, count, and count verification are based on the assumptions 1-6. Maximum, its verification, and median algorithms are based on assumptions 1-8.

We assume, any DB owner or the initiator provides additive shares of $m$ to servers for executing PSI, and the DB owners have only positive integers to compute the max. Since the current PSI maximum method uses modular operations (as will be clear in~\S\ref{subsec:psi_maximum}), we cannot handle floating-point values \emph{directly}. Nonetheless, we can find the maximum for a large class of practical situations, where the precision of decimal is limited, say $k>0$ digits by simply multiplying each number by $10^k$  and using the current PSI maximum algorithm. E.g., we can find the maximum over $\{0.5,$$ 8.2,8.02\}$ by computing the maximum over $\{50,820,802\}$. Designing a more general solution that does not require limited precision is non-trivial.

\medskip
\noindent\textbf{Parameters known to servers.} Servers know following parameters:
(\textit{i}) $m$, $\delta>m$, the generator $g$ of the cyclic (sub)group of order $\delta$ and $\eta^{\prime}=\alpha\times\eta$ and $\alpha>1$. Based on the group theory, $\eta-1$ should be divisible by $\delta$. Note, servers do not know $\eta$.
%
(\textit{ii}) A permutation function $\mathcal{PF}$, and recall that the same permutation function is also known to DB owners.
%
(\textit{iii}) Two permutation functions $\mathcal{PF}_{s1}$ and $\mathcal{PF}_{s2}$.
   %
(\textit{iv}) A common pseudo-random number generator $\mathcal{PRG}$ that generates random numbers between 1 and $\delta-1$; $\mathcal{PRG}$ is unknown to DB owners. PSI, sum, and average are based on the assumptions 1. Maximum, its verification, and median are based on the assumptions 1,2. Count and its verification are based on the assumptions 1,3. PSU algorithm is based on the assumptions 1,4.

}

\B
\section{Private Set Intersection Query}
\label{sec:Common Item Finding}
\B
This section, first, develops a method for finding PSI among $m>2$ different DB owners on an attribute $A_c$ (which is assumed to exist at all DB owners, \S\ref{subsec:Single Attribute Common Item Finding}) and presents a result verification method (\S\ref{subsec:psi verification}). Later in \S\ref{subsec:Bucketization-based PSI}, we present a method to execute PSI over multiple attributes and a method to reduce the communication cost of PSI.


\subsection{PSI Query Execution}
\label{subsec:Single Attribute Common Item Finding}


 \noindent\textbf{High-level idea.} 
 Each of $m>2$ DB owners uses a publicly known hash function to map distinct values of $A_c$ attribute in a table of at most $|\mathrm{Dom}(A_c)|$ cells, where $|\mathrm{Dom}(A_c)|$ is the size of the domain of $A_c$. Thus, if a value $a_j\in A_c$ exists at any DB owner, all DB owners must map $a_j$ to an identical cell of the table. All values of the table are outsourced in the form of additive shares to \emph{two non-communicating servers} $\mathcal{S}_\phi$, $\phi\in\{1,2\}$, that \emph{obliviously} find the common items/intersection and return shared output vector (of the same length as the length of the received shares from DB owners). Finally, each DB owner adds the results to know the final answer.

\smallskip
\noindent\textbf{{Construction}.} We create the following construction over elements of a group under addition and elements of a cyclic group under multiplication. We can select any cyclic group such that $\eta>m$.
\begin{equation}
\label{eq:common_item_finding_construction}
\begin{aligned}
  (x+y)\bmod \delta = 0, (g^{x}\times g^y)\bmod \eta = 1
  \end{aligned}
\end{equation}

Based on this construction, below, we explain PSI finding algorithm:

\medskip
\noindent\textbf{\textsc{Step} 1: {DB owners.}}
{\color{black} Each DB owner finds distinct values in an attribute ($A_c$, which exists at all DB owners, as per our assumption given in~\S\ref{sec:Assumptions Related to Parameters}) and executes the hash function on each value $a_i$ to create a table $\chi=\{x_1,x_2,\ldots, x_b\}$ of length $b=|\mathrm{Dom}(A_c)|$. The hash function maps the value $a_i\in A_{\mathit{c}}$ to one of the cells of $\chi$, such that the cell of $\chi$ corresponding to the value $a_i$ holds 1; otherwise 0.\footnote{\scriptsize We can also add any positive random number except 1 in case of 0 to prevent revealing data distribution based on background knowledge; see~\cite{full_version} for details.}} It is important that each cell must contain only a single one corresponding to the unique value of the attribute $A_c$, and note that if a value $a_i\in A_c$ exists at any DB owner, then one corresponding to $a_i$ is  placed at an identical cell of $\chi$ at the DB owner. The table at $\mathcal{DB}_j$ is denoted by $\chi_{j}$. Finally, $\mathcal{DB}_j$ creates additive secret-shares of each value of $\chi_j$ (\textit{i}.\textit{e}., additive secret-shares of either one or zero) and outsources the $\phi^{\mathit{th}}$, $\phi\in\{1,2\}$, share to the server $\mathcal{S}_\phi$. We use the notation $A(x_i)^\phi_j$ to refer to $\phi^{\mathit{th}}$ additive share of an $i^{\mathit{th}}$ element of $\chi_j$ of $\mathcal{DB}_j$. {\color{black} Recall that before the computation starts, the initiator informs the locations of servers to DB owners and vice versa (\S\ref{subsec:Entities}).}

\smallskip
\noindent\textbf{\textsc{Step} 2: {Servers.}}
Each server $\mathcal{S}_\phi$ ($\phi\in \{1,2\}$) holds the $\phi^{\mathit{th}}$ additive share of the table $\chi$ (denoted by $A(\chi)_j^{\phi}$) of $j^{\mathit{th}}$ ($1\leq j\leq m$) DB owners and executes Equation~\ref{eq:common_item_finding_server}:
\begin{equation}
\label{eq:common_item_finding_server}
\begin{aligned}
  \mathit{output}_i^{\mathcal{S}_\phi} \leftarrow  g^{((\oplus_{j=1}^{j=m} A(x_i)_j^{\phi}) \ominus A(m)^{\phi})} \bmod \eta^{\prime},  (1\leq i \leq b)
  \end{aligned}
\end{equation}
where $\oplus$ and $\ominus$ show the modular addition and modular subtraction operations, respectively. We used the symbols $\oplus$ and $\ominus$ to distinguish them from the normal addition and subtraction. Particularly, each server $\mathcal{S}_\phi$ performs the following operations:
(\textit{i}) modular addition (under $\delta$) of the $i^{\mathit{th}}$ additive secret-shares from all $m$ DB owners,
(\textit{ii}) modular subtraction (under $\delta$) of the result of the previous step from the additive share of $m$ (\textit{i}.\textit{e}., $A(m)^{\phi}$),
(\textit{iii}) exponentiation by $g$ to the power the result of the previous step and modulo by $\eta^{\prime}$, and
(\textit{iv}) sends all the computed $b$ results to the $m$ DB owners.

\smallskip
\noindent\textbf{\textsc{Step} 3: {DB owners.}} From two servers, DB owners receive two vectors, each of length $b$, and perform modular multiplication (under $\eta$) of outputs
$\mathit{output}_i^{\mathcal{S}_1}$ and $\mathit{output}_i^{\mathcal{S}_2}$, where $1\leq i\leq b$, \textit{i}.\textit{e}.,
\begin{equation}
\label{eq:common_item_finding_db_owner_final_processing}
\begin{aligned}
  \mathit{fop}_i\leftarrow (output_i^{\mathcal{S}_1} \times output_i^{\mathcal{S}_2})\bmod \eta
  \end{aligned}
\end{equation}
This step results in an output array of $b$ elements, which may contain any value. However, if an $i^{\mathit{th}}$ item of $\chi_j$ exists at all DB owners, then $\mathit{fop}_i$ must be one, 
since $\mathcal{S}_\phi$ have added additive shares of $m$ ones at the $i^{\mathit{th}}$ element and subtracted from additive share of $m$ that results in $(g^{0}\bmod \eta^{\prime})\bmod \eta=1$ at DB owner. Please see the correctness argument below after the example.

 \medskip
\noindent\textbf{Example~\ref{sec:Common Item Finding}.1.} Assume three DB owners: $\mathcal{DB}_1$, $\mathcal{DB}_2$, and $\mathcal{DB}_3$; see Tables~\ref{tab:table 1},~\ref{tab:table 2}, and~\ref{tab:table 3}. For answering a query to find the common disease that is treated by each hospital, DB owners create their tables $\chi$ as shown in the first column of Tables~\ref{tab:user 1},~\ref{tab:user 2}, and~\ref{tab:user 3}.
For example, in Table~\ref{tab:user 2}, $\langle 1,1,0\rangle$ corresponds to cancer, fever, and heart diseases, where 1 means that the disease is treated by the hospital. We select $\delta=5$, $\eta=11$, and $\eta^{\prime}=143$. Hence, the Abelian group under modulo addition contains $\{0,1,2,3,4\}$, and the cyclic (sub)group (with $g=3$) under modulo multiplication contains $\{1,3,4,5,9\}$. Assume additive shares of $m=3=(1+2)\bmod 5$.

\noindent\textit{\textsc{Step} 1: DB Owners.} DB owners generate additive shares as shown in the second and third columns of Tables~\ref{tab:user 1},~\ref{tab:user 2}, and~\ref{tab:user 3}, and outsource all values of the second and third columns to $\mathcal{S}_1$ and $\mathcal{S}_2$, respectively.

\begin{table}[!t]
\scriptsize
\centering
\begin{minipage}{0.3\linewidth}
\centering
\bgroup
\def\arraystretch{.9}
\BB
\centering
\begin{tabular}{|p{0.35cm}|p{0.5cm}|p{0.5cm}|}\hline
Value & Share~1 & Share~2 \\ \hline\hline
1 & 4 & -3   \\ \hline
0 & 2 & -2  \\ \hline
1 & 3 & -2  \\  \hline
\end{tabular}
\caption{$\mathcal{DB}_1$.}
\label{tab:user 1}
\egroup
\end{minipage}
\begin{minipage}{0.3\linewidth}
\centering
\bgroup
\def\arraystretch{0.9}
\BB
\centering
\begin{tabular}{|p{0.35cm}|p{0.5cm}|p{0.5cm}|}\hline
Value & Share~1 & Share~2 \\ \hline\hline
1 & 3 & -2   \\ \hline
1 & 4 & -3  \\ \hline
0 & 3 & -3  \\  \hline
\end{tabular}
\caption{$\mathcal{DB}_2$.}
\label{tab:user 2}
\egroup
\end{minipage}
\begin{minipage}{0.3\linewidth}
\centering
\bgroup
\def\arraystretch{0.9}
\BB
\centering
\begin{tabular}{|p{0.35cm}|p{0.5cm}|p{0.5cm}|}\hline
Value & Share~1 & Share~2 \\ \hline\hline
1 & 2 & -1   \\ \hline
0 & 3 & -3  \\ \hline
1 & 4 & -3  \\  \hline
\end{tabular}
\caption{$\mathcal{DB}_3$.}
\label{tab:user 3}
\egroup
\end{minipage}
\BBB\BBB\BB
\end{table}

\noindent\textit{\textsc{Step} 2: {Servers.}} The server $\mathcal{S}_1$ will return the three values 27, 27, 81, by executing the following computation, to all three DB owners:

\centerline{$3^{((((4+3+2)\bmod 5)-1)\bmod 5)} \bmod 143=27$}
\centerline{$3^{((((2+4+3)\bmod 5)-1)\bmod 5)} \bmod 143=27$}
\centerline{$3^{((((3+3+4)\bmod 5)-1)\bmod 5)} \bmod 143=81$}
\noindent The server $\mathcal{S}_2$ will return values 9, 1, and 1 to all three DB owners:

\centerline{$3^{((((-3-2-1)\bmod 5)-2)\bmod 5)}\bmod 143=9$}
\centerline{$3^{((((-2-3-3)\bmod 5)-2)\bmod 5)}\bmod 143=1$}
\centerline{$3^{((((-2-3-3)\bmod 5)-2)\bmod 5)}\bmod 143=1$ }

\noindent\textit{\textsc{Step} 3: {DB owners.}} The DB owner obtains a vector $\langle 1, 5, 4\rangle$, by executing the following computation (see below). From the vector $\langle 1, 5, 4\rangle$, DB owners learn that cancer is a common disease treated by all three hospitals. 
However, the DB owner does not learn anything more than this; note that in the output vector, the values 5 and 4 correspond to zero. For instance, $\mathcal{DB}_1$, \textit{i}.\textit{e}., hospital 1, cannot learn whether fever and heart diseases are treated by hospital 2, 3, or not.

\centerline{$(27\times 9)\bmod 11=1 \quad (27\times 1)\bmod 11=5 \quad (81\times 1)\bmod 11=4$}


\medskip
\noindent\textbf{Correctness.}
When we plug Equation~\ref{eq:common_item_finding_server} into Equation~\ref{eq:common_item_finding_db_owner_final_processing}, we obtain:

$
\begin{aligned}
  \mathit{fop}_i &= (g^{(\oplus^{j=m}_{j=1}A(x_i)_j^1)\ominus A(m)^1}\times g^{(\oplus^{j=m}_{j=1}A(x_i)_j^2)\ominus A(m)^2}\bmod {\eta^{\prime}})\bmod \eta\\
        &= (g^{(\oplus^{j=m}_{j=1}(x_i)_j-m})\bmod{\eta^{\prime}})\bmod \eta
\end{aligned}
$

We utilize modular identity, \textit{i}.\textit{e}., $(x\bmod{\alpha\eta})\bmod{\eta} = x\bmod{\eta}$; thus, $\mathit{fop}_i=g^{(\sum^{j=m}_{j=1}(x_i)_j-m})\bmod{\eta}$.
Only when $\sum^{j=m}_{j=1}(x_i)_j=m$, the result of above expression is one; otherwise, a nonzero number.

\medskip
\noindent\textbf{Information leakage discussion.}
We need to prevent information leakage at the server and at the DB owners.
\begin{enumerate}[nolistsep,noitemsep,leftmargin=0.01in]
  \item 
  \emph{Server perspective.} Servers only know the parameters $\langle g,\delta, \eta^{\prime}\rangle$ and may utilize the relations between $g$ and $\eta$ to guess $\eta$ from $\eta^{\prime}$. However, it will not give any meaningful information to servers, since the DB owner sends the elements of $\chi$ in additive shared form, and since servers do not communicate with each other, they cannot obtain the cleartext values of $\chi$. Also, an identical operation is executed on all shares of $m$ DB owners. Hence, access-patterns are hidden from servers, preventing them to distinguish between any two values based on access-patterns. Also, the output of queries is in shared form and contains an identical number of bits as inputs. Thus, based on the output size, servers cannot know whether the value is common among DB owners or not.

  \item 
  \emph{DB owner perspective.} When all DB owners do not have one at the $i^{\mathit{th}}$ position of $\chi$, we need to inform DB owners that there is no common value and not to reveal that how many DB owners do not have one at the $i^{\mathit{th}}$ position. Note that the DB owner can learn this information, if they know $g$ and $\alpha$, since based on these values, they can compute what the servers have computed. However, unawareness of $g$ and $\alpha$ makes it impossible to guess the number of DB owners that do not have one at the $i^{\mathit{th}}$ position of $\chi$. We can formally prove it as follows:

\smallskip
\noindent
\textbf{Lemma.}
A DB owner cannot deduce how many other DB owners do not have one at the $i^{\mathit{th}}$ position of $\chi$ without knowing $g$.

\noindent
\textbf{Proof.} According to the precondition, $g$ is a generator of a cyclic group of order $\delta$, where $\delta$ is a prime number. Thus, $\mathcal{C}=\{g^0, g, g^2, \ldots, g^{\delta-1}\}$ represents all items in the cyclic group. Assume that the output of Equation~\ref{eq:common_item_finding_db_owner_final_processing} is a number other than one, say $\beta$. Thus, we have
$\beta=g^{x-m} \bmod \eta$, where $x$ represents the number of one at the $i^{\mathit{th}}$ position of $\chi_j$, $1\leq j\leq m$. When DB owners wish to know $x$, they must compute $\log_g \beta$. To solve it, they need to know $g$. 
Note that based on the characteristic of the cyclic group, there are
less than $\delta-1$ generators of $\mathcal{C}$ and co-prime to $\delta$. Thus, $g^2,\ldots, g^{\delta-1}$ may also be generators of the cyclic group. However, DB owners cannot distinguish which generator is used by the servers. Thus, DB owners cannot deduce the value of $x$, except knowing that $x\in[0, m-1]$.\footnote{{\scriptsize Consider $i^{\mathit{th}}$,
$j^{\mathit{th}}$, and $k^{\mathit{th}}$ values of $\chi_1=\{1,0,0\}$,  $\chi_2=\{0,1,0\}$, $\chi_3=\{1,1,1\}$. Here, after \textsc{Step} 3, DB owners will learn three random numbers, such that the first two random numbers will be identical. Based on this, DB owner can only know that the sum of $i^{\mathit{th}}$ and $j^{\mathit{th}}$ position of $\chi$ is identical. However, it will not reveal how many positions have 0 or 1 at $i^{\mathit{th}}$ or $j^{\mathit{th}}$ positions.}}$\blacksquare$
\end{enumerate}

\subsection{PSI Result Verification}
\label{subsec:psi verification}
A malicious adversary or a hardware/software bug may result in the following situations,  during computing PSI:
(\textit{i}) skip processing the $i^{\mathit{th}}$ additive shares of all DB owners,
(\textit{ii}) replacing the result of the $i^{\mathit{th}}$ additive shares by the computed result for $j^{\mathit{th}}$ share,
(\textit{iii}) {\color{black} injecting fake values}, or
(\textit{iv}) falsifying the verification method.
Thus, this section provides a method for verifying the result of PSI. 

\smallskip
\noindent
\textbf{High-level idea.}  
Let $g$ be a generator of a cyclic group under modulo multiplicative $\eta$ operation, and $\eta^{\prime}=\alpha\times \eta$, $\alpha>1$. Thus, $(g^x \bmod \eta) \times (g^{-x}\bmod \eta)=1$, and the idea of PSI verification lies in this equation. Recall, in PSI (\S\ref{subsec:Single Attribute Common Item Finding}), we used $(g^x \bmod \eta)$, whose value 1 shows that the item exists at all DB owners. Now, we will use the term $(g^{-x}\bmod\eta)$ for verification. Specifically, if the servers has performed their computations correctly, then Equation~\ref{eq:PSI_verification_main_eq} must hold to be true:
\begin{equation}
\label{eq:PSI_verification_main_eq}
\begin{aligned}
((g^{(\oplus_{j=1}^{j=m} A(x_i)_j^{\phi})-A(m)^{\phi}}\bmod \eta^{\prime})  \times
(g^{\oplus_{j=1}^{j=m} \overline{A(x_i)_j^{\phi}}} \bmod \eta^{\prime}))\bmod \eta  =1
\end{aligned}
\end{equation}
where $m$ is the number of DB owners, $x_j$ is either 1 or 0 (as described in \S\ref{subsec:Single Attribute Common Item Finding}), and $\overline{x_j}$ is the complement value of $x_j$. Below, we describe the steps executed at the servers and DB owners.

\noindent
\textbf{\textsc{Step} 1: DB owners.}
On distinct values of an attribute $A_\mathit{c}$ of their relations, $\mathcal{DB}_j$ executes a hash function to create the table $\chi_j$ that contains $b=|\mathrm{Dom}(A_c)|$ values (either 0 or 1). Also, $\mathcal{DB}_j$ creates a table $\overline{\chi_j}$ containing $b$ values, such that $i^{\mathit{th}}$ value of $\overline{\chi_j}$ must be the complement of $i^{\mathit{th}}$ value of $\chi_j$. Then, $\mathcal{DB}_j$ permutes the values of $\overline{\chi_j}$ using a permutation function $\mathcal{PF}_{db1}$ (known to all DB owners \emph{only}) and creates additive shares of each value of $\chi_j$ and $\overline{\chi_j}$, prior to outsourcing to servers. Reason of using $\mathcal{PF}_{db1}$ will be clear soon.

\noindent\textbf{\textsc{Step} 2: {Servers.}}
Each server $\mathcal{S}_\phi$ holds the $\phi^{\mathit{th}}$ additive share of $\chi$ (denoted by $A(\chi)_j^{\phi}$) and $\overline{\chi}$ (denoted by $A(\overline{\chi})_j^{\phi}$) of $j^{\mathit{th}}$ DB owner and executes the following operation:
\begin{equation}
\label{eq:PSI_verification_server_eq1}
\begin{aligned}
\mathit{output}_i^{\mathcal{S}_\phi} \leftarrow g^{((\oplus_{j=1}^{j=m} A(x_i)_j^{\phi}) \ominus A(m)^{\phi})} \bmod \eta^{\prime}, (1\leq i \leq b)
\end{aligned}
\end{equation}
\BBB
\begin{equation}
\label{eq:PSI_verification_server_eq2}
\begin{aligned}
\mathit{Vout}_i^{\mathcal{S}_\phi} \leftarrow g^{((\oplus_{j=1}^{j=m} A(\overline{x_i})_j^{\phi}))} \bmod \eta^{\prime}, (1\leq i \leq b)
\end{aligned}
\end{equation}

Equation~\ref{eq:PSI_verification_server_eq1} is identical to Equation~\ref{eq:common_item_finding_server} (in \S\ref{subsec:Single Attribute Common Item Finding}) and finds the common item at the server. In Equation~\ref{eq:PSI_verification_server_eq2}, each server $\mathcal{S}_\phi$ performs following operations:
(\textit{i}) modular addition (under $\delta$) of the $i^{\mathit{th}}$ additive shares of $\overline{\chi}$ from $m$ DB owners,
(\textit{ii}) exponentiation by $g$ to the power the result of previous step, under modulo $\eta^{\prime}$; and
(\textit{iii}) sends computed results $\mathit{output}^{\mathcal{S}_\phi}[]$ and $\mathit{Vout}^{\mathcal{S}_\phi}[]$ to DB owners.

\noindent\textbf{\textsc{Step} 3: {DB owners.}} From two servers, DB owners receive $\mathit{output}^{\mathcal{S}_\phi}[]$ and $\mathit{Vout}^{\mathcal{S}_\phi}[]$ (each of length $b$), permute back the values of $\mathit{Vout}^{\mathcal{S}_\phi}[]$ (using the reverse permutation function, since they used $\mathcal{PF}_{db1}$ on $\overline{\chi}$, which results in $\mathit{Vout}^{\mathcal{S}_\phi}[]$ at servers) to obtain $\mathit{pvout}^{\mathcal{S}_\phi}[]$, and execute the following:
\begin{equation}
\label{eq:PSI_verification_user_checking1}
\begin{aligned}
r_1\leftarrow \mathit{output}_i^{\mathcal{S}_1}\times \mathit{output}_i^{\mathcal{S}_2} \bmod \eta
\end{aligned}
\end{equation}
\BB
\begin{equation}
\label{eq:PSI_verification_user_checking2}
\begin{aligned}
r_2\leftarrow \mathit{pvout}_i^{\mathcal{S}_1}\times \mathit{pvout}_i^{\mathcal{S}_2} \bmod \eta
\end{aligned}
\end{equation}
\BBB
\begin{equation}
\label{eq:PSI_verification_user_checking3}
\begin{aligned}
r_1\times r_2 \bmod \eta \: ? \: 1
\end{aligned}
\end{equation}
If the DB owner finds the output of $r_1 \times r_2$ equals one for all $b$ values, it shows that the servers executed the computation correctly.

\medskip

\noindent\textbf{Example~\ref{subsec:psi verification}.1.} We verify PSI results of Example~\ref{subsec:Single Attribute Common Item Finding}.1. Suppose $\delta=5$, $\eta=11$, and $\eta^{\prime}=143$, as assumed in Example~\ref{subsec:Single Attribute Common Item Finding}.1.

\noindent\textit{\textsc{Step} 1: {DB owners.}} DB owners find the reverse of $\chi$ (as shown in the first column of Tables~\ref{tab:verify user 1},~\ref{tab:verify user 2}, and~\ref{tab:verify user 3}) and generate additive shares; see the second and third columns of Tables~\ref{tab:verify user 1},~\ref{tab:verify user 2}, and~\ref{tab:verify user 3}. Note that here for simplicity, we do not permute the values or shares. 

\begin{table}[!t]
\B
\scriptsize
\centering
\begin{minipage}{0.3\linewidth}
\centering
\bgroup
\def\arraystretch{.97}
\centering
\begin{tabular}{|p{0.35cm}|p{0.5cm}|p{0.5cm}|}\hline
$\overline{\textnormal{Value}}$ & Share~1 & Share~2 \\ \hline\hline
0 & 2 & -2   \\ \hline
1 & 0 & 1  \\ \hline
0 & 1 & -1  \\  \hline
\end{tabular}
\caption{$\mathcal{DB}_1$.}
\label{tab:verify user 1}
\egroup
\end{minipage}
\begin{minipage}{0.3\linewidth}
\centering
\bgroup
\def\arraystretch{0.97}
\centering
\begin{tabular}{|p{0.35cm}|p{0.5cm}|p{0.5cm}|}\hline
$\overline{\textnormal{Value}}$ & Share~1 & Share~2 \\ \hline\hline
0 & 2 & -2   \\ \hline
0 & 3 & -3  \\ \hline
1 & 4 & -3  \\  \hline
\end{tabular}
\caption{$\mathcal{DB}_2$.}
\label{tab:verify user 2}
\egroup
\end{minipage}
\begin{minipage}{0.3\linewidth}
\centering
\bgroup
\def\arraystretch{0.97}

\centering
\begin{tabular}{|p{0.35cm}|p{0.5cm}|p{0.5cm}|}\hline
$\overline{\textnormal{Value}}$ & Share~1 & Share~2 \\ \hline\hline
0 & 4 & -4   \\ \hline
1 & 1 & 0  \\ \hline
0 & 1 & -1  \\  \hline
\end{tabular}

\caption{$\mathcal{DB}_3$.}
\label{tab:verify user 3}
\egroup
\end{minipage}
\BBB\BBB\BBB
\end{table}

\noindent\textit{\textsc{Step} 2: {Servers.}} The server $\mathcal{S}_1$ will return the three values 27, 81, 3, by executing the following computation, to all three DB owners:

 \centerline{$3^{((2+2+4)\bmod 5)} \bmod 143=27$}
 \centerline{$3^{((0+3+1)\bmod 5)} \bmod 143=81$}
 \centerline{$3^{((1+4+1)\bmod 5)} \bmod 143=3$}

\noindent $\mathcal{S}_2$ will return three values 7, 27, and 1 to all three DB owners:

\centerline{$3^{((-2-2-4)\bmod 5)}\bmod 143=9$}
\centerline{$3^{((1-3+0)\bmod 5)}\bmod 143=27$}
\centerline{$3^{((-1-3-1)\bmod 5)}\bmod 143=1$ }

\noindent\textit{\textsc{Step} 3: {DB owners.}} The DB owner obtains a vector $\langle 1, 9, 8\rangle$, by executing the following computation:

\centerline{$(27\times 9)\bmod 11=1 \quad (81\times 27)\bmod 11=9 \quad (3\times 1)\bmod 11=3$}

Now, the DB owner executes the following to verify PSI results:
$1\times 1 \bmod 11=1$,
$5\times 9 \bmod 11=1$, and
$4\times 3 \bmod 11=1$,  where 1, 5, 4 are final outputs at DB owner in Example~\ref{subsec:Single Attribute Common Item Finding}.1.
The output 1 indicates that servers executed the computation correctly. $\blacksquare$

\medskip
\noindent
\textbf{Correctness.} First, we need to argue that the processing at servers works correctly.
Assume that the DB owner does not implement $\mathcal{PF}_{db1}$ on elements of $\overline{\chi}$, and computation at servers is executed in cleartext. Thus, on the values of $\chi$, servers add $i^{\mathit{th}}$ value of each $\chi_j=\{x_i\}$ ($1\leq j\leq m$, $1\leq i\leq b$) and subtract the results from $m$. It will result in a number, say $a\in\{-m+1, 0\}$. 
On the other hand, servers add $i^{\mathit{th}}$ values $\overline{\chi}_j$, and it will result in a number, say $b\in \{0,m\}$, \textit{i}.\textit{e}., the number of ones at DB owners at the $i^{\mathit{th}}$ position of $\chi$. To hide the value of $a$ and $b$ from servers, they execute operations on additive shares of $\chi$ and $\overline{\chi}$, and take a modulus exponent (\textit{i}.\textit{e}., $r_1\leftarrow g^{a}$ and $r_2\leftarrow g^b$) to hide $a$ and $b$ from DB owners. Since $a=-b$ or $a=b=0$, $r_1\times r_2\bmod \eta=1$, and this shows that the server executed the correct operation.

Now, we show why the verification method will detect any abnormal computation executed by servers. Note that servers may skip processing all/some values of $\chi$ and $\overline{\chi}$. For example, servers may process only $x_1\in \chi$, $\overline{x_1}\in \overline{\chi}$, and send the results corresponding to $x_1$, $\overline{x_1}$ as the results of all remaining $b-1$ values. Such a malicious operation of servers will provide legal proof (\textit{i}.\textit{e}., $r_1\times r_2\bmod \eta=1$) at DB owners that servers executed the computation correctly, (since values of $\overline{\chi}$ was not permuted). Thus, we used permutation over the values of ~$\overline{\chi}$ and/or additive shares of $\overline{\chi}$. Now, to break the verification method and to produce $r_1\times r_2\bmod \eta=1$ for an $i^{\mathit{th}}$ value of $\chi$, servers need to find the correct value in $\overline{\chi}$ corresponding to an $i^{\mathit{th}}$ value of $\chi$ (among the randomly permuted shares). Hence, the removal of any results from the output will be detected.

{\color{black} Now, we show that the verification method can detect fake data insertion by servers. For a server $\mathcal{S}_1$ to successfully inject a fake tuple (\textit{i}.\textit{e}., undetected during verification), it should know the correct position of some element in both $A(\chi)_j^1$ and $A(\overline{\chi})_j^1$. Since $A(\overline{\chi})_j^1$ is a permuted vector of size $b=|\mathrm{Dom}(A_c)|$, the probability of finding the correct element in $A(\overline{\chi})_j^1$ corresponding to an element of $A(\chi)_j^1$ will be $1/b^2$.
E.g., in our experiments, the domain size is
5M (or 20M) values, making the above probability infinitesimal ($< 10^{-13}$).\footnote{\scriptsize {\color{black}If the domain size is small, we can increase its size by adding fake values to bind the probability of adversary being able to inject fake data.}}

\smallskip
\noindent\textbf{Additional security.} We implemented $\mathcal{PF}_{\mathit{db1}}$ on the elements of $\overline{\chi}$. We can, further, permute additive shares of both  ${\chi}$ and $\overline{\chi}$ using different permutation functions, to make it impossible for both servers to find the position of a value in $A(\chi)_j^{\phi}$ and $A(\overline{\chi})_j^{\phi}$, $\phi\in\{1,2\}$. Thus, servers cannot break the verification method, and any malicious activities will be detected by DB owners.
}

\smallskip
\noindent
\textbf{Information leakages discussion.}
The verification method will not reveal any non-desired information to servers/DB owners, and
arguments follow the similar way as for PSI computation in \S\ref{subsec:Single Attribute Common Item Finding}.

\section{Aggregation Operation over PSI}
\label{sec:Extending PSI}

\textsc{Prism} supports both summary and exemplar aggregations. Below, we describe how \textsc{Prism} implements sum~\S\ref{subsec:psi_Sum Query}, average~\S\ref{subsec:psi_average}, maximum~\S\ref{subsec:psi_maximum}, median~\S\ref{subsec:psi_Median Query} and count operations~\S\ref{subsec:psi_Count Query}. Also, in our discussion below, we consider set-based operation PSI on a single attribute $A_c$. \S\ref{subsec:Bucketization-based PSI} extends the discussions to support PSI over multiple attributes and over a large-size domain. \emph{Correctness and information leakage discussion of the following methods with their verification approaches are given in the full version~\cite{full_version}}.

\subsection{PSI Sum Query}
\label{subsec:psi_Sum Query}

A PSI sum query computes the sum of values over an attribute corresponding to common items in another attribute; see example given in \S\ref{sec:Private Set Operations}. 
This section develops a method based on additive, as well as, multiplicative shares, where additive shares find common items over an attribute $A_c$ and multiplicative shares (SSS) finds the sum of shares of an attribute $A_x$ corresponding to the common items in $A_c$. This method contains the following steps:

\smallskip
\noindent\textbf{\textsc{Step} 1: {DB owners.}} $\mathcal{DB}_j$ creates their $\chi_j$ table over the distinct values of $A_c$ attribute by following \textsc{Step} 1 of PSI (\S\ref{sec:Common Item Finding}). Here, $\chi_j= \{\langle x_{i1},x_{i2} \rangle \}$ ($1\leq i\leq b$ and $b=|\mathrm{Dom}(A_c)|$), \textit{i}.\textit{e}., the $i^{\mathit{th}}$ cell of $\chi_j$ contains a pair of values, $\langle x_{i1}, x_{i2}\rangle$, where
(\textit{i}) $x_{i1}=1$, if a value $a_i\in A_c$ is mapped to the $i^{\mathit{th}}$ cell, otherwise, 0; and
(\textit{ii}) $x_{i2}$ contains the sum of values of $A_x$ attribute corresponding to $a_i$; otherwise, 0. $\mathcal{DB}_j$ creates additive shares of $x_{i1}$ (denoted by $A(x_{i1})^\phi_j$, $\phi=\{1,2\}$) and sends to servers $\mathcal{S}_1$ and $\mathcal{S}_2$. $\mathcal{DB}_j$ also creates SSS of $x_{i2}$ (denoted by $S(x_{i2})^{\phi=\{1,2,3\}}$) and sends to  servers $\mathcal{S}_1$, $\mathcal{S}_2$, and $\mathcal{S}_3$.

\smallskip
\noindent\textbf{\textsc{Step} 2: {Servers.}}
Servers $\mathcal{S}_1$ and $\mathcal{S}_2$ find common items using additive shares by implementing
Equation~\ref{eq:common_item_finding_server} and send all computed $b$ results to all DB owners. Since the result is in additive shared form, it cannot be multiplied to SSS. Thus, {\color{black} servers send the output of PSI to \emph{one of the DB owners selected randomly} and wait to receive multiplicative shares corresponding to common items. The reason of randomly selecting only one DB owner is just to reduce the communication overhead of sending/receiving additive/multiplicative shares, and it does not impact the security. Note that all DB owners can receive the PSI outputs and generate multiplicative shares.}


\smallskip
\noindent\textbf{\textsc{Step} 3: {DB owners.}} On receiving $b$ values, the  DB owner
finds the common items by executing Equation~\ref{eq:common_item_finding_db_owner_final_processing} and generates a vector of length $b$ having 1 or 0 only, where 0 is obtained by replacing random values of $\mathit{fop}$. Finally, DB owner creates three SSS of each $b$ values, denoted by $S(z_i)^\phi$, $\phi=\{1,2,3\}$, and sends to three servers.

\smallskip
\noindent\textbf{\textsc{Step} 4: {Servers.}} Servers $\mathcal{S}_\phi$, $\phi=\{1,2,3\}$, execute the following:
\begin{equation}
\label{eq:psi_sum_server}
\begin{aligned}
  \mathit{sum}_i^{\mathcal{S}_\phi} \leftarrow \textstyle \sum_{j=1}^{j=m} S(x_{i2})_j^\phi \times S(z_i)^\phi, 1\leq i \leq b
  \end{aligned}
\end{equation}
Each server multiplies $S(z_i)^\phi$ by $S(x_{i2})_j^\phi$ of each DB owner, adds the results, and sends them to all DB owners.

\smallskip
\noindent\textbf{\textsc{Step} 5: {DB owners.}} From three servers, DB owners receive three vectors, each of length $b$, and perform Lagrange interpolation on each $i^{\mathit{th}}$ value of the three vectors to obtain the final sum of the value in $A_x$ corresponding to the common items in $A_c$. 

\subsection{PSI Average Query}
\label{subsec:psi_average}
A PSI average query on cost column corresponding to the common disease in Tables~\ref{tab:table 1}-\ref{tab:table 3} returns $\{$Cancer, 280$\}$. PSI average query works in a similar way as PSI sum query. In short, $\mathcal{DB}_j$ creates $\chi_j= \{\langle x_{i1},x_{i2}, x_{i3} \rangle \}$ ($1\leq i\leq b$, $b=|\mathrm{Dom}(A_c)|$), and $x_{i1}, x_{i2}$ are identical to the values we created in \textsc{Step} 1 of PSI sum (\S\ref{subsec:psi_Sum Query}). The new value $x_{i3}$ contains the number of tuples at $\mathcal{DB}_j$ corresponding to $x_{i1}$. E.g., in case of Table~\ref{tab:table 1}, one of the values of $\chi_1$ will be $\{\langle$Cancer, 300, 2$\rangle \}$, \textit{i}.\textit{e}., Table~\ref{tab:table 1} has two tuples corresponding to Cancer and cost 300. All values $x_{i3}$ are outsourced in multiplicative share form. Then, we follow \textsc{Steps} 2 and 3 of PSI sum. In \textsc{Step 4}, servers multiply the received $i^{\mathit{th}}$ SSS values corresponding to the common value to $x_{i2}, x_{i3}$ and add the values. Finally, in \textsc{Step} 5, DB owners interpolate vectors corresponding to all $b$ values of $x_{i2}, x_{i3}$ and find the average by dividing the values appropriately.

\subsection{PSI Maximum Query}
\label{subsec:psi_maximum}
This section develops a method for finding the maximum value in an attribute $A_x$ corresponding to the common values in $A_c$ attribute; refer to \S\ref{sec:Private Set Operations} for PSI maximum example.
Here, our objective is to prevent the adversarial server from learning: (\textit{i}) the actual maximum values outsourced by each DB owner, (\textit{ii}) what is the maximum value among DB owners and which DB owners have the maximum value. We allow all the DB owners to know the maximum value and/or the identity of the DB owner(s) having the maximum value. \textcolor[rgb]{1.00,0.00,1.00}{We use pink color to highlight the part that is used to reveal the identity of DB owners having maximum to distinguish which part of the algorithm can be avoided based on the security requirements.}

In this method, \textbf{\emph{each DB owner uses polynomial}} $\boldsymbol{\mathcal{F}(x)}$ given by the initiator (see \S\ref{sec:Assumptions Related to Parameters} to find how we created $\mathcal{F}(x)$). We use $\mathcal{F}(x)$ to generate values at different DB owners in an order-preserving manner by executing the following \textsc{Step}~3 and Equation~\ref{eq:qmax1_db_owner_step_1_round 1}.

{\color{black} The method contains at most three rounds, where the first round finds the common values in an attribute $A_c$ by using \textsc{Step}s 1-3,  the second round finds the maximum value in an attribute $A_x$ corresponding to common items in $A_c$ using \textsc{Step}s 4-5a, the last round finds DB owners who have the maximum value using \textsc{Step}s 5b-7.} Note that \emph{\textbf{the third round is not always required}}, if (\textit{i}) we do not want to reveal identity of the DB owner having the maximum value, or (\textit{ii}) values in $A_x$ column across all DB owners are unique. 



\medskip
\noindent{\color{black}\textbf{{\textsc{Step} 1 at DB owner and \textsc{Step} 2 at servers.}} }These two steps are identical to \textsc{Step} 1 and \textsc{Step} 2 of PSI query execution method (\S\ref{sec:Common Item Finding}).

\smallskip
\noindent{\color{black}\textbf{{\textsc{Step} 3: DB owner.}}} On the received outputs (of \textsc{Step} 2) from servers, DB owners find the common item in the attribute $A_c$, as in \textsc{Step} 3 of PSI query execution method (\S\ref{sec:Common Item Finding}). Now, to find the maximum value in the attribute $A_{\mathit{x}}$ corresponding to the common item in  $A_c$, DB owners proceeds as follows:

For simplicity, we assume that there is only one common item, say $y^{\mathit{th}}$ item. $\mathcal{DB}_i$ finds the maximum, say $\mathcal{M}_{iy}$, in the attribute $A_{\mathit{x}}$ of their relation corresponding to the common item $y$. Note that since we assume only one common element, we refer to the maximum element $\mathcal{M}_{iy}$ by $\mathcal{M}_{i}$.  $\mathcal{DB}_i$ executes Equation~\ref{eq:qmax1_db_owner_step_1_round 1} to produce values at DB owners in an  order-preserving manner:
\begin{equation}
\label{eq:qmax1_db_owner_step_1_round 1}
\begin{aligned}
v_i \leftarrow \mathcal{F}(\mathcal{M}_i)+r_i
\end{aligned}
\end{equation}
$\mathcal{DB}_i$ implements the polynomial $\mathcal{F}()$ on $\mathcal{M}_i$ and adds a random number $r_i$ (selected in a range between 0 and $\mathcal{M}_i^m$), and it produces a value $v_i$. Finally, $\mathcal{DB}_i$ creates additive shares of $v_i$ (denoted by $A(v)_i^{\phi}$) and sends them to servers $\mathcal{S}_\phi$, $\phi=\{1,2\}$. Note that even {if $k\geq 2$ DB owners have the same maximum value $\mathcal{M}_i$, by this step, the value $v$ will be different at those DB owners, with a high probability, $1-\frac{1}{(\mathcal{M}_i)^{(k-1)m}}$, (depending on the range of $r_i$). Also, if any two numbers $\mathcal{M}_i<\mathcal{M}_j$, then$\mathcal{F}(\mathcal{M}_i)+r_i<\mathcal{F}(\mathcal{M}_j)$ will hold.} 

\smallskip
\noindent{\color{black}\textbf{\textsc{Step} 4: Servers.}} Each server $\mathcal{S}_\phi$ executes the following operation:
\begin{equation*}
\label{eq:qmax1_server_step_1_round 1}
\begin{aligned}
\mathit{input}^{\mathcal{S}_\phi}[i] \leftarrow A(v)_i^{\phi}, 1\leq i\leq m; \mathit{output}^{\mathcal{S}_\phi}[] \leftarrow \mathcal{PF}(\mathit{input}^{\mathcal{S}_\phi}[])
\end{aligned}
\end{equation*}
$\mathcal{S}_\phi$ collects additive shares from each DB owner and places them in an array (denoted by $\mathit{input}^{\mathcal{S}_\phi}[]$), on which $\mathcal{S}_\phi$ executes the permutation function $\mathcal{PF}$. Then, $\mathcal{S}_\phi$ sends the output the permutation function  $\mathit{output}^{\mathcal{S}_\phi}[]$ to the announcer $\mathcal{S}_a$ that does the following:
\begin{equation}
\label{eq:s_a_adds_shares}
\begin{aligned}
\mathit{foutput}^{\mathcal{S}_a}[i]\leftarrow \mathit{output}^{\mathcal{S}_1}[i] + \mathit{output}^{\mathcal{S}_2}[i], 1\leq i\leq m
\end{aligned}
\end{equation}
\begin{equation}
\label{eq:s_a_find_max_and_index}
\begin{aligned}
\mathit{max},\textcolor[rgb]{1.00,0.00,1.00}{\mathit{index}}\leftarrow \mathit{FindMax}(\mathit{foutput}^{\mathcal{S}_a}[])
\end{aligned}
\end{equation}
$\mathcal{S}_a$ adds the $i^{\mathit{th}}$ outputs received from $\mathcal{S}_1$ and $\mathcal{S}_2$, and compares all those numbers to find the maximum number (denoted by $\mathit{max}$). \textcolor[rgb]{1.00,0.00,1.00}{Also, $\mathcal{S}_a$ produces the index position (denoted by $\mathit{index}$) corresponding to the maximum number in $\mathit{foutput}^{\mathcal{S}_a}[]$.} Finally, $\mathcal{S}_a$ creates additive secret-shares of $\mathit{max}$ (denoted by $A(\mathit{max}^{\mathcal{S}_\phi})$, $\phi\in\{1,2\}$), as well as, of \textcolor[rgb]{1.00,0.00,1.00}{$\mathit{index}$ (denoted by $A(\mathit{index})^{\mathcal{S}_\phi}$)}, and sends them to $\mathcal{S}_\phi$ that forwards such additive shares to DB owners. Note, \textcolor[rgb]{1.00,0.00,1.00}{if the \emph{protocol does not require to reveal the identity} of the DB owner having the maximum value, $\mathcal{S}_a$ does not send additive shares of $\mathit{index}$}.

\smallskip
\noindent{\color{black}\textbf{{\textsc{Step} 5a: DB owner.}}} Now, the DB owners' task is to find the maximum value and/or the identity of the DB owner who has the maximum value. To do so, each DB owner performs the following:
\begin{equation}
\label{eq:max_verification_user_checking1}
\begin{aligned}
\mathsf{max}\leftarrow A(\mathit{max})^{\mathcal{S}_1} + A(\mathit{max})^{\mathcal{S}_2}
\end{aligned}
\end{equation}
\begin{equation}
\label{eq:max_verification_user_checking2}
\begin{aligned}
\textcolor[rgb]{1.00,0.00,1.00}{\mathsf{index}\leftarrow A(\mathit{index})^{\mathcal{S}_1}+A(\mathit{index})^{\mathcal{S}_2}, \mathit{pos}\leftarrow \mathcal{RPF}(\mathsf{index})
}\end{aligned}
\end{equation}
\textcolor[rgb]{1.00,0.00,1.00}{The DB owner finds the identity of the DB owner having the max value by adding the additive shares and by implementing reverse permutation function $\mathcal{RPF}$. ($\mathcal{RPF}$ works since $\mathcal{PF}$ is known to DB owners and servers; see Assumptions given in~\S\ref{sec:Assumptions Related to Parameters})}.
To find the max value, they implement $\mathcal{F}(z)$ and $\mathcal{F}(z+1)$ and evaluate $\mathcal{F}(z)\leq \mathsf{max}<\mathcal{F}(z+1)$, where $z\in \{1,2,\ldots\}$.\footnote{{\scriptsize To reduce the computation cost, we can select number $z$ using binary search method.}} If this condition holds to be true, then $z$ is the max value, and if $z=\mathcal{M}_i$, then $\mathcal{DB}_i$ knows that he/she holds the max value. Obviously, if  $\mathcal{DB}_i$ does not hold the max value, then
$\mathcal{M}_i< \mathcal{F}(\mathcal{M}_i)+r_i < \mathcal{F}(\mathcal{M}_i+1)\leq \mathcal{F}(z)\leq \mathsf{max}$.

\smallskip
\noindent\textcolor[rgb]{1.00,0.00,1.00}{\textbf{{\color{black}{\textsc{Step} 5b: DB owner.}} } By the end of \textsc{Step} 5a, the DB owners know the max value and the identity of the DB owner having the same max value, due to $\mathit{pos}$. But, if there are more than one DB owners having the max value, the other DB owners cannot learn about it. The reason is: the server $\mathcal{S}_a$ can find only the max value, while, recall that, if more than one DB owners have the same max value, say $\mathcal{M}$, they produce a different value, due to using different random numbers in \textsc{Step} 3 (Equation~\ref{eq:qmax1_db_owner_step_1_round 1}). Thus, we need to execute this step 5b to know all DB owners having the max value.}
\textcolor[rgb]{1.00,0.00,1.00}{
After comparing its max values against $\mathsf{max}$, $\mathcal{DB}_i$ knows whether it possesses the maxi value or not. Depending on this, $\mathcal{DB}_i$ generates a value $\alpha_i=0$ or $\alpha_i=1$, creates additive shares of $\alpha_i$, and sends to $\mathcal{S}_\phi$, $\phi\in \{1,2\}$.}

\smallskip
\noindent\textcolor[rgb]{1.00,0.00,1.00}{\textbf{\color{black}{{\textsc{Step} 6: Servers.}}} Server $\mathcal{S}_\phi$ allocates the received additive shares to a vector, denoted by $\mathit{fpos}$, and sends the vector $\mathit{fpos}$ to all DB owners, \textit{i}.\textit{e}., $\mathit{fpos}^{\mathcal{S}_\phi}[i] \leftarrow A(\alpha)_i^{\mathcal{S}_\phi}, 1\leq i\leq m$.}

\noindent\textcolor[rgb]{1.00,0.00,1.00}{\textbf{\color{black}{{\textsc{Step} 7: DB owner.}}} Each DB owner adds the received additive shares to obtain the vector $\mathsf{fpos}[]$.
\begin{equation}
\label{eq:PSI_verification_db ower step 6}
\begin{aligned}
\mathsf{fpos}[i]\leftarrow \mathit{fpos}^{\mathcal{S}_1}[i]+\mathit{fpos}^{\mathcal{S}_2}[i], 1\leq i\leq m
\end{aligned}
\end{equation}
By $\mathsf{fpos}[]$, DB owners discover which DB owners have the maximum value, since, recall that in \textsc{Step} 5, $\mathcal{DB}_i$ that satisfies the condition ($\mathcal{F}(\mathcal{M}_i)\leq \mathsf{max}<\mathcal{F}(\mathcal{M}_i+1)$) requests $\mathcal{S}_\phi$ to place additive share of 1 at $\mathit{fpos}^{\mathcal{S}_\phi}[i]$.
}

\noindent\textbf{Example~\ref{subsec:psi_maximum}.1.}  Refer to Tables~\ref{tab:table 1}-\ref{tab:table 3}, and consider that all hospitals wish to find the maximum age of a patient corresponding to the common disease and which hospitals treat such patients. Assume $\eta=5003$ and that all hospitals know cancer as the common disease.


In \textsc{Step} 3, all hospitals, \textit{i}.\textit{e}., DB owners, find their maximum values in the attribute \texttt{Age} corresponding to common disease and implement  $\mathcal{F}(x)=x^4+x^3+x^2+x+1$, sent by the initiator.

\centerline{$\mathcal{F}(6)=1555+216=1771 = (5000-3229) \bmod 5003$}
\centerline{$\mathcal{F}(8)=4681+1  =4682 = (5500-818)  \bmod 5003$}
\centerline{$\mathcal{F}(8)=4681+319=5000 = (2500+2500) \bmod 5003$}
Further, they add random numbers ($216, 1, 319$) and create additive shares, which are outsourced to $\mathcal{S}_1$ and $\mathcal{S}_2$.
In \textsc{Step} 4, $\mathcal{S}_1$ holds $\langle5000,5500,2500\rangle$, permutes them, and sends to $\mathcal{S}_a$. $\mathcal{S}_2$ holds $\langle-3229,-818,2500\rangle$, permutes them, and sends to $\mathcal{S}_a$.

$\mathcal{S}_a$ obtains $\langle4682,5000,1771\rangle$ by adding the received shares from $\mathcal{S}_1$, $\mathcal{S}_2$, and finds 5000 as the max value and `Hospital 2' to which this value belongs. Finally, $\mathcal{S}_a$ creates additive shares of $5000=(4000+1000)\bmod 5003$,
\textcolor[rgb]{1.00,0.00,1.00}{additive shares of the identity of the DB owner: $2=(200-198)\bmod 5003$,} and sends to DB owners via $\mathcal{S}_\phi$.

In \textsc{Step} 5a, all hospitals will know the maximum value as 5000 (with random value added) \textcolor[rgb]{1.00,0.00,1.00}{and identity of the DB owner as 2 on which they implement the reverse permutation function to obtain the correct identity as `Hospital 3'}. Then, `Hospital 1' knows that they do not hold the maximum, since $\mathcal{F}(6)+216<\mathcal{F}(7)<5000$. `Hospital 2' knows that they hold the maximum, since $\mathcal{F}(8)<5000<\mathcal{F}(9)$. Also, `Hospital 3' knows that they hold the maximum.

\noindent
\textcolor[rgb]{1.00,0.00,1.00}{
\noindent To know which hospitals have the maximum value, in \textsc{Step} 5b, Hospitals 1, 2, 3' create additive shares of 0, 1, 1, respectively, as:  $0=(200-200)\bmod 5003$, $1=(300-299)\bmod 5003$, and $1=(200-199)\bmod 5003$, and send to $\mathcal{S}_1$ and $\mathcal{S}_2$. Finally, in \textsc{Step} 6, $\mathcal{S}_1$ and $\mathcal{S}_2$ send $\langle200,300,200\rangle$ and $\langle-200,-299,-199\rangle$ to all hospitals. In \textsc{Step} 7,
hospitals add received shares, resulting in $\langle0,1,1\rangle$. It shows that `Hospitals 2, 3' have the maximum value 8.} $\blacksquare$

\subsection{PSI Median Query}
\label{subsec:psi_Median Query}

A PSI median query over cost column corresponding to disease column over Tables~\ref{tab:table 1}-\ref{tab:table 3} returns $\{\langle$Cancer, 300$\rangle\}$ (here, we first added the cost of treatment per disease at each DB owner). 
For PSI median, we extend the method of finding max by executing all steps as specified in~\S\ref{subsec:psi_maximum} with an additional process in \textsc{Step} 2. Particularly, $\mathcal{S}_a$ in \textsc{Step} 2 of \S\ref{subsec:psi_maximum} after adding shares, sorts them, and finds the median value. If number of DB owners is odd (even), then $\mathcal{S}_a$ finds the middle (two middle) values in the sorted shares.

\subsection{PSI Count Query}
\label{subsec:psi_Count Query}

We extend PSI method (\S\ref{sec:Common Item Finding}) to only reveal the count of common items among DB owners (\textit{i}.\textit{e}., the cardinality of the common item), instead of revealing common items.
Recall that servers $\mathcal{S}_\phi$ know a permutation function $\mathcal{PF}_{s1}$ that is not known to DB owners. The idea behind this is to find the common items over $\chi$ and to permute the final output at servers before sending the vector (of additive share form) to DB owners. Thus, when DB owners perform computation on the vector received from servers to know the final output, the position of one in the vector will not reveal common items, while the count of one will reveal the cardinality of the common items.
Thus, PSI count method follows all steps of PSI as described in \S\ref{subsec:Single Attribute Common Item Finding} with an addition of permutation function execution by servers before sending the output to DB owners. 

\begin{figure}[!t]
  \centering
  \includegraphics[scale=0.3]{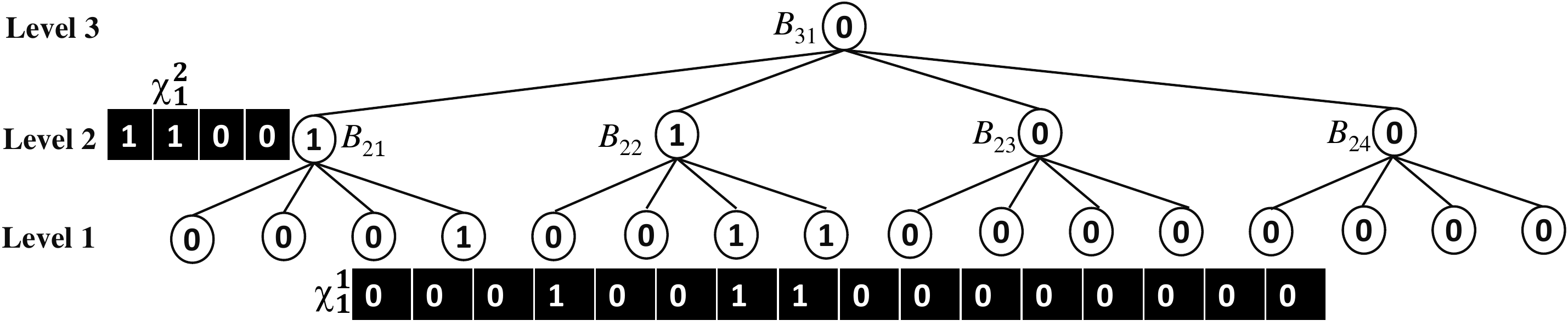}
 \BBB\BBB\BBB
  \caption{Bucket tree for 16 values.}
  \label{fig:bucket_tree}
  \BBB\B
\end{figure}

\subsection{Extending PSI over Multiple Attributes}
\label{subsec:Bucketization-based PSI}
In the previous sections, we explained PSI over a single attribute (or a set). We can trivially extend it to multiple attributes (or multisets). Particularly, such a query can be express in SQL as follows:

{\scriptsize
\centerline{
\texttt{SELECT $A_c$, $A_x$ FROM $db_1$
INTERSECT
$\ldots$
INTERSECT
SELECT $A_c$, $A_x$ FROM $\mathit{db}_m$}}}

Recall that in PSI finding method~ \S\ref{subsec:Single Attribute Common Item Finding}, $\mathcal{DB}_j$ sends additive shares of a table $\chi_j$ of length $b=|\mathrm{Dom}(A_c)|$, where $A_c$ was the attributes on which we executed PSI. Now, we can extend this method by creating a table $\chi_j$ of length $b=|\Pi_{i>0}\mathrm{Dom}(A_i)|$, where $A_i$ are attributes on which we want to execute PSI. However, as the domain size and the number of attributes increase, such a method incurs the communication overhead. Thus, {\color{black}to apply the PSI method over a large (and real) domain size, as well as, to reduce the communication overhead, we provide a method, named as bucketization-based PSI.}

\smallskip
\noindent
\textbf{Optimization: bucketization-based PSI.} Before going to steps of this method, let us consider the following example:


\noindent
\textbf{Example~\ref{subsec:Bucketization-based PSI}.1.} Consider two attributes $A$ with $|\mathrm{Dom}(A)|=8$ and $B$ with $|\mathrm{Dom}(B)|=2$. Thus, DB owners can create $\chi_j$ of 16 cells.  Assume that there are two DB owners: $\mathcal{DB}_1$ with $\chi_1$ whose only positions $4,7,8$ have one; and $\mathcal{DB}_2$ with $\chi_2$ whose only positions $1,6,8$ have one. Thus, each DB owner sends/receives a vector of length 16 from each server.
Now, to reduce communication, we create buckets over the cell of~$\chi$ and build a tree, called \emph{bucket-tree}, of depth $\log_\kappa |\chi|$, where $\kappa$ is the number of the maximum number of child nodes that a node can have. Bucket-tree in created in a bottom-up manner, by non-overlap grouping of $\kappa$ nodes. For each level of bucket-tree a hash table is created (similar to $\chi$). Notation $\chi^i_j$ denotes this table for $i^{\mathit{th}}$ level of bucket-tree at~$\mathcal{DB}_j$,~and $\chi^i_j[k]=1$, if $k^{\mathit{th}}$ node at the $i^{\mathit{th}}$ level has 1.

Figure~\ref{fig:bucket_tree} shows bucket-tree for $\mathcal{DB}_j$, $|\chi|=16$, and $\kappa=4$, with appropriate one and zero in $\chi^i_1$. Note that the second level shows four nodes $B_{21},B_{22},B_{23},B_{24}$ corresponding to $1-4$, $5-8$, $9-12$, and $13-16$. Since $\mathcal{DB}_1$ has one at $4,7,8$ leaf nodes, we obtain $\chi^2_1=\langle 1,1,0,0\rangle$, \textit{i}.\textit{e}., $B_{21}=1,B_{22}=1,B_{23}=0,B_{24}=0$. Here, $B_{21}=1$, since  one of its child nodes has one. Now, when computing PSI, $\mathcal{DB}_j$ starts the computation shown in \textsc{Step} 2 of \S\ref{subsec:Single Attribute Common Item Finding} over the specified $i^{\mathit{th}}$ levels' $\chi^i_j$. The computation is continued only for the child nodes, whose parent nodes resulted in one in \textsc{Step} 3 of~\S\ref{subsec:Single Attribute Common Item Finding}.

For example, in Figure~\ref{fig:bucket_tree}, $\mathcal{DB}_j$ can execute PSI for $\chi^2_j$ and know that the only desired bucket nodes are $B_{21}$ and $B_{22}$ that contain common items. Thus, in the next round, they execute PSI over the first eight items of $\chi^1_j$, \textit{i}.\textit{e}., child nodes of $B_{21}$ and $B_{22}$. Hence, while we use two communication rounds, DB owners/servers send 4+8=12 numbers instead of 16 numbers. $\blacksquare$

Bucketization-based PSI has the following steps:

\noindent\textbf{\textsc{Step 1a}: DB owner.} Build the tree as specified in Example~\ref{subsec:Bucketization-based PSI}.1.

\noindent\textbf{\textsc{Step 1b}: DB owner.} Outsource additive shares of $i^{\mathit{th}}$ level's $\chi^i_j$.

\noindent\textbf{\textsc{Step} 2: Servers.} Servers compute PSI using \textsc{Step} 2 of \S\ref{subsec:Single Attribute Common Item Finding} over $\chi^i_j$ ($1\leq j\leq m$) and provide answers to DB owners.

\noindent\textbf{\textsc{Step} 3: DB owner.} $\mathcal{DB}_j$ computes results to find the common items in $\chi^i_j$ and discards all non-common values of $\chi^i_j$ and their child nodes. $\mathcal{DB}_j$ requests servers to execute the above \textsc{Step 2} for $\chi^{i-1}_j$ that has values corresponding to all non-discarded nodes of $(i-1)^{\mathit{th}}$ level node. \noindent\textbf{Note:} The role of DB owners in traversing the tree (\textit{i}.\textit{e}., the above \textsc{Step} 3) can be eliminated by involving $\mathcal{S}_a$.

\smallskip
\noindent
\textbf{Open problem.}
In bucketization, we perform PSI at layers of the tree to eliminate ranges where corresponding child nodes have zero.
However, if the data is dense (\textit{i}.\textit{e}., data covers most of the domain values), then bucketization-based PSI may incur overhead, since all nodes
in the tree may correspond to one, leading to PSI execution on all those nodes 
including leaf nodes. In contrast, if the data is sparse (\textit{i}.\textit{e}., the domain is
much larger than the data, as is the case of the domain to be a cartesian product of
domains of two or more attributes), then higher-level nodes in the tree may have 0, leading to eliminate ranges of the domain on which PSI is performed. Developing an optimal bucketization strategy that minimizes PSI execution is an interesting open problem.

\begin{figure*}[!t]
\BBB\B
		\begin{center}
		\centering
   \includegraphics[scale=0.5]{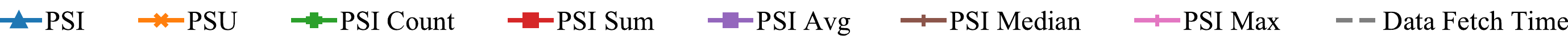}
    \B
		\end{center}
		\BBB\BBB
	\end{figure*}

\begin{figure}[!t]
		\begin{center}
			\begin{minipage}{.49\linewidth}
				\centering
			\includegraphics[scale=0.49]{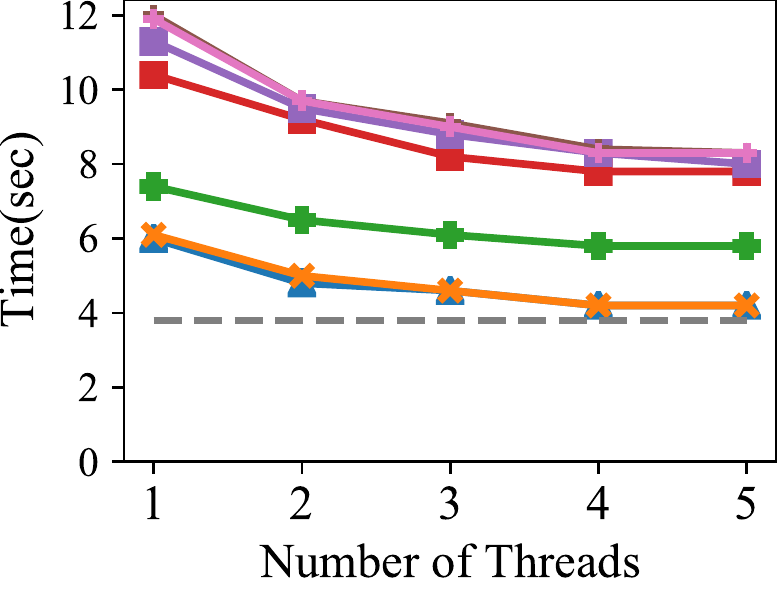}\BBB
				\subcaption{5M OK domain size (1-5M).}
				\label{fig:5M rows parallelism}
			\end{minipage}
			\begin{minipage}{.49\linewidth}
				\centering
				\includegraphics[scale=0.49]{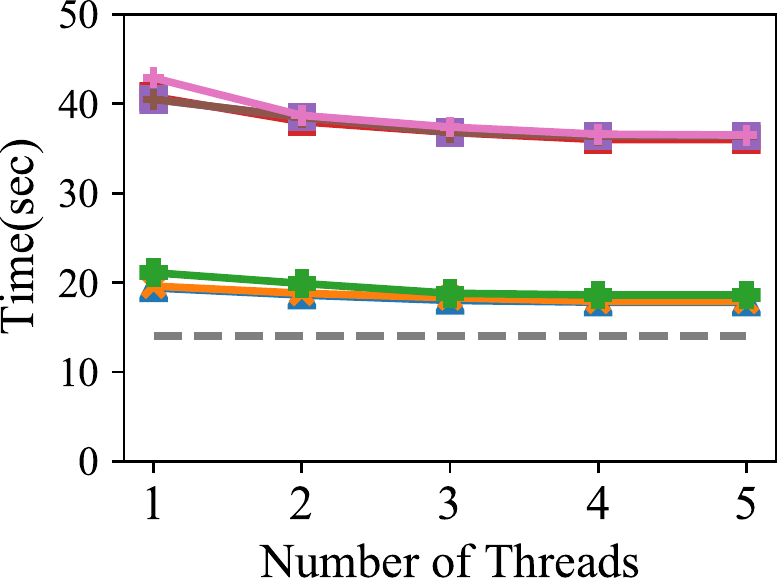}\BBB
				\subcaption{20M OK domain size (1-20M).}
				\label{fig:20M rows parallelism}
			\end{minipage}			
			  \end{center}
		\BBB\BBB
		\caption{Exp 1. \textsc{Prism} performance on multi-threaded implementation at AWS.}
		\label{fig:impact of parallelism}
		\BBB
	\end{figure}

\section{Private Set Union (PSU) Query}
\label{sec:Generalized Union Query}
This section develops a method for finding the union of values among $m>1$ different DB owners over an attribute $A_c$.
%

%

\noindent\textbf{High-level idea.} Likewise PSI method (as given in \S\ref{sec:Common Item Finding}), each DB owner uses a publicly known hash function to map distinct values of $A_c$ attribute in a table of cells at most $|\mathrm{Dom}(A_c)|$, where $|\mathrm{Dom}(A_c)|$ refers to the size of the domain of $A_c$, and outsources each element of the table in additive share form to \emph{two servers} $\mathcal{S}_\phi$, $\phi\in\{1,2\}$. $\mathcal{S}_\phi$ computes the union obliviously, thereby DB owners will receive a vector of length $|\mathrm{Dom}(A_c)|$ having either 0 or 1 of additive shared form. After adding the share for an $i^{\mathit{th}}$ element, DB owners only know whether the element is in the union or not; nothing else.

%

\noindent\textbf{\textsc{Step} 1: {DB owner.}} This step is identical to \textsc{Step} 1 of PSI (\S\ref{subsec:Single Attribute Common Item Finding}).

\noindent\textbf{\textsc{Step} 2: {Server.}}
Server $\mathcal{S}_\phi$ holds the $\phi^{\mathit{th}}$ additive share of the table $\chi$ of $m$ DB owners and executes the following operation:
\begin{equation}
\label{eq:guion_server}
\begin{aligned}
  \mathit{rand}[] &\leftarrow \mathcal{PRG}(\mathit{seed})\\
  \mathit{output}_i^{\mathcal{S}_\phi} &\leftarrow ((\textstyle\sum_{j=1}^{j=m} A(x_i)_j^{\phi}) \times \mathit{rand}[i])\bmod \delta
  \end{aligned}
\end{equation}
Server $\mathcal{S}_\phi$:
(\textit{i}) generates $b$ pseudorandom numbers that are between 1 and $\delta-1$,
(\textit{ii}) performs (arithmetic) addition of the $i^{\mathit{th}}$ additive secret-shares from all DB owners,
(\textit{iii}) multiplies the resultant of the previous step with $i^{\mathit{th}}$ pseudorandom number and then takes modulo, and
(\textit{iv}) sends $b$ results to all DB owners.

\noindent\textbf{\textsc{Step} 3: {DB owner.}} On receiving
two vectors, each of length $b$, from two servers, DB owners execute modular addition over $i^{\mathit{th}}$ shares of both vectors to get the final answer (Equation~\ref{eq:guion_db_owner}). It results in either zero or a random number, where zero shows that $i^{\mathit{th}}$ element of $\chi$ is not present at any DB owner, while a random number shows $i^{\mathit{th}}$ element of $\chi$ is present at one of the DB owners.
\begin{equation}
\label{eq:guion_db_owner}
\begin{aligned}
   \mathit{fop}_i &\leftarrow
   (\mathit{output}_i^{\mathcal{S}_1}+\mathit{output}_i^{\mathcal{S}_2})\bmod \delta
  \end{aligned}
\end{equation}

\section{Experimental Evaluation}
\label{sec:Experiments}

This section evaluates the scalability of \textsc{Prism} on different-sized datasets and a different number of DB owners. Also, we compare \textsc{Prism} against other MPC-based systems. We used a 16GB RAM machine with 4 cores for each of the DB owners and {\color{black}three AWS servers of 32GB RAM, 3.5GHz Intel Xeon CPU with 16 cores to store shares. Communication between DB owners and servers were done using the \texttt{scp} protocol, and $\eta$, $\delta$ were 227, 113, respectively.}

\bgroup
\def\arraystretch{.9}
\begin{table}[!t]
\small
\centering
\begin{tabular}{|p{0.3cm}|p{0.3cm}|p{0.3cm}|p{0.3cm}|p{0.3cm}|p{0.4cm}|p{0.4cm}|p{0.4cm}|p{0.4cm}|p{0.4cm}|p{0.8cm}|}\hline

\multicolumn{5}{|c|}{Real data column} & \multicolumn{5}{|c|}{For verification}& Average\\\hline
OK & PK & LN & SK & DT & vOK & vPK & vLN & vSK & vDT &aOK  \\ \hline
\end{tabular}
\caption{Table structure created by \textsc{Prism}.}
\label{tab:exp_data_table_db_owner}
\BBB\BBB\BB
\end{table}
\egroup

\BB
\subsection{\textbf{\textsc{{\large Prism}}} Evaluation}
\noindent\textbf{Dataset generation.} We used five columns (Orderkey (OK), Partkey (PK), Linenumber (LN), Suppkey(SK), and Discount (DT)) of LineItem table of TPC-H benchmark. {\color{black} We experimented with domain sizes (\textit{i}.\textit{e}., the number of values) of 5M and 20M for the \textbf{\emph{OK column}} on which we executed PSI and PSU.} Further, we selected at most \textbf{\emph{50 DB owners}}. To our knowledge, this is the first such experiment of multi-owner large datasets. OK column is used for PSI/PSU, and other columns were used for aggregation operations. To generate secret-shared dataset, each DB owner maintained a LineItem table containing at most 5M (20M) OK values. To outsource the database, each DB owner did the following:

\begin{enumerate}[noitemsep,nolistsep,leftmargin=0.01in]

\item Created a table of 11 columns, as shown in Table~\ref{tab:exp_data_table_db_owner}, in which the first five columns contain the secret-shared data of LineItem table, the next five columns contain the corresponding verification data, and the last column (aOK) was used for computing the average. All verification column names are prefixed with the character `v.'

\item First column of Table~\ref{tab:exp_data_table_db_owner} was created over OK column of LineItem table (using \textsc{Step} 1 of \S\ref{subsec:Single Attribute Common Item Finding}) for executing PSI/PSU over OK. vOK column was created to verify PSI results (using \textsc{Step} 1 of~\S\ref{subsec:psi verification}).

\item Columns PK and vPK were created using the following command: \texttt{select OK, sum(PK) from LineItem group by OK}. Other columns $\langle$LN, SK, DT, vLN, vSK, vDT$\rangle$ were created by using similar SQL commands.
Column aOK was created using the following command:  \texttt{select count(*) from LineItem group by OK}.

\item Finally, permuted all values of all verification columns and create additive shares of $\langle$OK and vOK$\rangle$, as well as, multiplicative shares of all remaining columns.
\end{enumerate}

\smallskip
{\color{black} \noindent\textbf{Share generation time.}  The time to generate two additive shares and three multiplicative shares of the respective first five columns of Table~\ref{tab:exp_data_table_db_owner} in the case of 5M (or 20M) OK domain size was 121s (or 548s). The time for creating each additional column for verification took 20s (or 90s) in the case of 5M (or 20M) domain values.}

	\begin{figure}[!t]
		\begin{center}
			\begin{minipage}{.49\linewidth}
				\centering
				\includegraphics[scale=0.49]{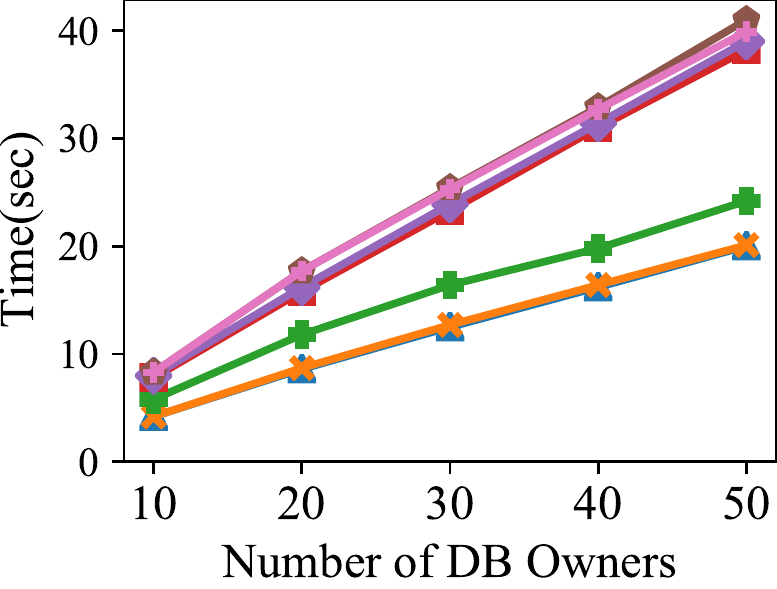}\B
				\subcaption{5M OK domain size (1-5M).}
				\label{fig:5M rows parallelism increasing DB owner}
			\end{minipage}
			\begin{minipage}{.49\linewidth}
				\centering
				\includegraphics[scale=0.49]{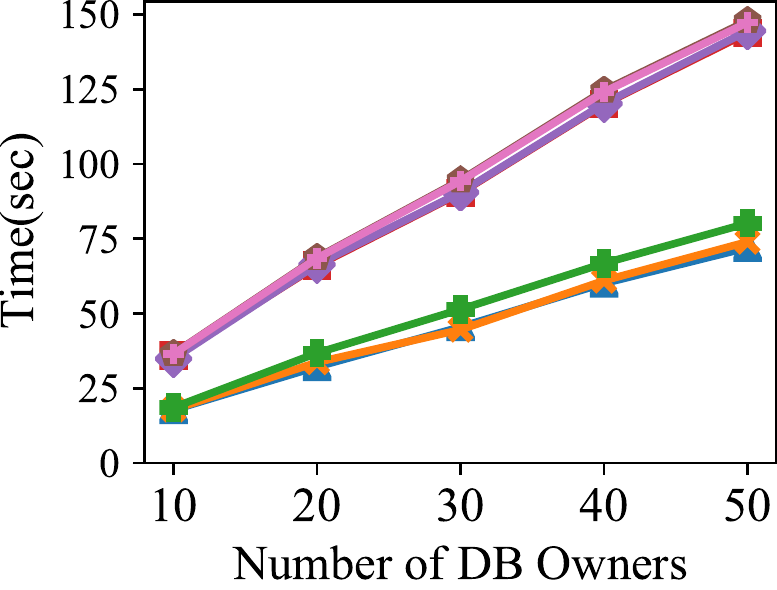}\B
				\subcaption{20M OK domain size (1-20M).}
				\label{fig:20M rows parallelism increasing DB owner}
			\end{minipage}			
			  \end{center}
		\BBB\BBB
		\caption{Exp 2. \textsc{Prism} dealing with multiple DB owners.}
		\label{fig:Prism dealing with multiple DB owners}
		\BBB\B
	\end{figure}

\smallskip
\noindent\textbf{Exp 1. \textsc{Prism} performance on multi-threaded implementation at AWS.}
Since identical computations are executed on each row of the table, we exploit multiple CPU cores by writing \textsc{Prism}'s the parallel implementation that divides rows into multiple blocks with each thread processing a single block. We increased the number of threads from 1 to 5; see Figure~\ref{fig:impact of parallelism}, while fixing DB owners to 10. Increasing threads more than 5 did not provide speed-up, since reading/writing of data quickly becomes the bottleneck as the number of threads increase. Observe that the data fetch time from the database remains (almost) identical; see Figure~\ref{fig:impact of parallelism}. 

\noindent\underline{\textit{PSI and PSU queries}.} Figure~\ref{fig:impact of parallelism} shows the time taken by PSI/PSU over OK column. Observe that as the number of values in OK column increases (from 5M to 20M), the time increases (almost) linearly from 4s to 18s, respectively.

\noindent\underline{\textit{Aggregation queries over PSI}.} We executed PSI count, average, sum, maximum, and median queries; see Figure~\ref{fig:impact of parallelism}. Observe that the processing time of PSI count is almost the same as that of PSI, since it involves only one round of computation in which we permute the output of PSI. In contrast, other aggregation operations (sum, average, maximum, and median) incur almost twice processing cost at servers, since they involve computing PSI over OK column in the first round and, then, computing aggregation in the second round. For this experiment, we computed the sum only over DT column and maximum/median over PK column. Table~\ref{tab:multi-column aggregation} shows the impact of computing sum and maximum over multiple attributes (from 1 to 4). As we increase the number of attributes, the computation of respective aggregation operations also increases, due to additional addition/multiplication/modulo operations on additional attributes.

\bgroup
\def\arraystretch{.9}
\begin{table}[!h]
\BB
\footnotesize
\centering
\begin{tabular}{|l|l|l|l|l|l|l|l|l|}\hline

\multirow{2}{*}{\textbf{Data size}} & \multicolumn{4}{|c|}{\textbf{Sum over different attributes}} & \multicolumn{4}{|c|}{\textbf{Max over different attributes}}\\\cline{2-9}
& \textbf{1} & \textbf{2} & \textbf{3} & \textbf{4} & \textbf{1} & \textbf{2} & \textbf{3} & \textbf{4}   \\ \hline
5M  & 8.2 & 12.1 & 15.9 & 20.4 & 10 & 14.6 & 19 & 23.5 \\ \hline
20M & 33.4 & 48.6 & 63.5 & 81.9 & 36.6 & 53.3 & 70 & 87.4 \\ \hline
\end{tabular}
\caption{Exp 1. Multi-column aggregation query performance (time in seconds).}
\label{tab:multi-column aggregation}
\BBB\BBB\BBB
\end{table}
\egroup

\begin{table*}[!t]
\BBB\BB
\scriptsize
\centering
\begin{tabular}{|p{3cm}|l|p{.7cm}|p{.8cm}|p{.7cm}|p{1cm}|p{.6cm}|p{1cm}|p{1.15cm}|p{1.15cm}|p{1.15cm}|p{1.2cm}|}\hline


\textbf{Papers} & \cite{DBLP:conf/ccs/Kerschbaum12} \& \cite{DBLP:conf/ic2e/LiuNZGH14} & \cite{DBLP:journals/tcc/QiuLSLW18} & \cite{DBLP:journals/tdsc/AbadiTMD19}   &  \cite{DBLP:conf/fc/AbadiTD16} & \cite{DBLP:conf/fc/KamaraM0S14} &
\cite{DBLP:conf/sac/Kerschbaum12} &
Jana~\cite{DBLP:journals/iacr/ArcherBLKNPSW18}${\dag}$  &
SMCQL~\cite{SMCQL} & Sharemind~\cite{Sharemind} &
Conclave~\cite{DBLP:conf/eurosys/VolgushevSGVLB19}${\ddag}$ &{\color{red} \textbf{\textsc{Prism}}}
\\\hline\hline

\textbf{Operations supported}  & PSI  & PSI & PSI &  PSI & PSI & PSI & PSI, PSU, aggregation & PSI via join \& aggregation & PSI via join \& aggregation & PSI via join \& aggregation  &  {\color{red}\textbf{PSI, PSU, aggregation}} \\
\hline

\textbf{Verification Support} & $\times$ &  $\times$ & $\times$ & $\checkmark$ & $\checkmark$ & $\times$ & $\times$ & $\times$ & $\times$ & $\times$ & {\color{red}$\boldsymbol{\checkmark}$} \\
\hline

\textbf{Scalability based on experiments reported (dataset size \& time)}  & N/A  & 32768  ($\approx$50 \textbf{m}) & 1 million ($\approx$2 \textbf{h}) & 32768 ($\approx$16 \textbf{m}) & 1 billion ($\approx$10 \textbf{m}) & 1000 ($\approx$9 \textbf{m}) & 1 million ($\approx$1 \textbf{h}) & $>$23 million ($\approx$23 \textbf{h}) & 30000 ($>$2 \textbf{h}) & 4 million  (8 \textbf{m}) &  {\color{red}\textbf{20 million}} {\color{red} \textbf{(At most 8 s)}}\\\hline

\textbf{Communication among servers}  & N/A  & N/A & N/A & N/A & N/A & N/A & \textbf{Yes} $\ast$ &  \textbf{Yes} $\ast$ &  \textbf{Yes} $\ast$ &  \textbf{Yes} $\ast$ & {\color{red} \textbf{No}} \\\hline

\textbf{Computational Complexity } &  $\mathcal{O}(n^m)$  &    $\mathcal{O}(\alpha mn)$ & $\mathcal{O}(n^m)$ & $\mathcal{O}(mn^2)$ & $\mathcal{O}(mn)$ $\ddag\ddag$ & $\mathcal{O}(n^m)$ & $\mathcal{O}(n^m)$ &  N/A $\ast$  & $\mathcal{O}(n^m)$ & N/A $\ast$ & {\color{red}$\boldsymbol{\mathcal{O}(mX)}$} \\\hline

\end{tabular}

\caption{Comparison of existing \emph{cloud-based} techniques against \textsc{Prism}.{\textnormal{{\scriptsize { Notes. (\textit{i}) The \textbf{scalability numbers are taken from the respective papers.} (\textit{ii}) Results of Sharemind~\cite{Sharemind} are taken from Conclave~\cite{DBLP:conf/eurosys/VolgushevSGVLB19} experimental comparison. (\textit{ii}) $\#$DB owners were in each paper was reported two; thus, we executed \textsc{Prism} for two DB owners for this table. (\textit{iv}) Only Jana, SMCQL, Sharemind, and Conclave provide identical security like \textsc{Prism}. (\textit{v}) \textbf{h}: hours. \textbf{m}: minutes. \textbf{s}: seconds. $\dag$: We setup Jana for two DB owners each with 1M values in our experiments. $\ddag$: Conclave~\cite{DBLP:conf/eurosys/VolgushevSGVLB19} uses a trusted party. 
\textbf{Yes}: Requires communication among servers. \textbf{No}: No communication among servers. $\ast$: Based on MPC-based systems. $\ast\ast$: N/A because executing operation in cleartext or at the trusted party. $m$: \#DB owners. $n$: DB size. $X$: domain size. $\ddag\ddag$: A insecure technique that reveals the size of the intersection, and hence fast. $\alpha$: The cost of Bilinear Map pairing technique. }}} }}
\label{tab:prev_papers_table}
\BBB\BBB\BB
\end{table*}

\noindent
\noindent\textbf{Exp 2. Impact of the number of DB owners.} \textsc{Prism}  deals with multiple DB owners; thus,
we investigated the impact of DB owners by selecting 10, 20, 30, 40, 50 DB owners, for two different domain sizes of OK column.
Figure~\ref{fig:Prism dealing with multiple DB owners} shows the server processing time for PSI, PSU, and aggregation over PSI. Observe that as the number of DB owners increases, the computation time at the server increases linearly, due to the linearly increasing number of addition/multiplication/modulo operations; \textit{e}.\textit{g}., on 5M OK values, PSI processing took 4.2s, 8.6s, 12.5s, 16.2s, and 20s in the case of 10, 20, 30, 40, 50 DB owners. 

\begin{figure}[!t]
\BBB
		\begin{center}
\begin{minipage}{.45\linewidth}	
\scriptsize
\centering
\vspace*{0.3cm}
\begin{tabular}{|l|l|l|}\hline

\textbf{Data Size} & \textbf{5M} & \textbf{20M} \\

\hline
\textbf{PSI} & 1.3 & 4.8 \\
\hline
\textbf{Count} & 1.7 & 5.4 \\
\hline
\textbf{Sum} & 3.1 & 10.3 \\
\hline
\textbf{Avg} & 3.2 & 10.3 \\
\hline
\textbf{Max} & 2.8 & 9.5 \\
\hline
\textbf{PSU} & 1.3 & 4.8 \\
\hline
\end{tabular}
\captionof{table}{Exp 3. DB owner processing time in result construction (in seconds).}
\label{tab:exp_db_owner_processing}
\BBB\BBB
\end{minipage}%
\hspace*{.4cm}
			\begin{minipage}{.45\linewidth}
				\centering
\includegraphics[scale=0.46]{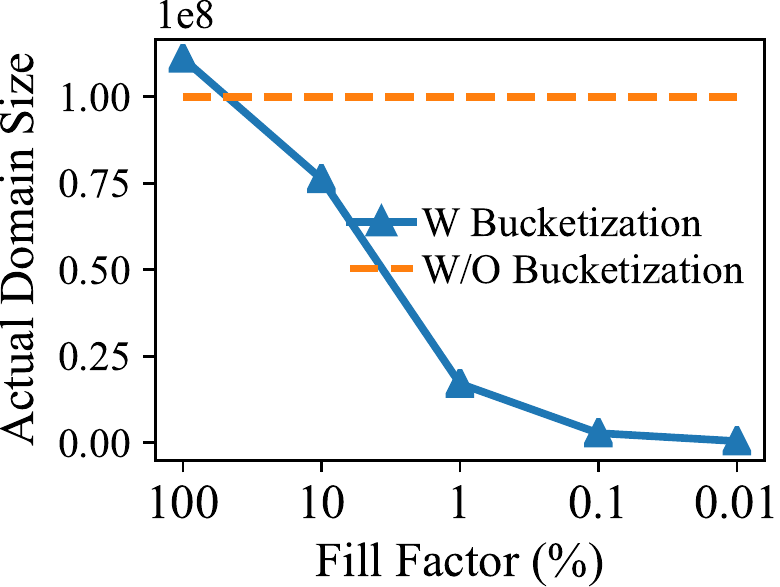}\BBB\BB
\caption{Exp 4. Impact of bucketization.}
			\label{fig:bucketization}
			\end{minipage}%

			  \end{center}
		\BBB\BBB

\end{figure}

\smallskip
\noindent
\noindent\textbf{Exp 3. DB owner processing time in result construction.} In \textsc{Prism}, DB owners perform computation on additive or multiplicative shares. Table~\ref{tab:exp_db_owner_processing} shows the processing time at a DB owner over 5M and 20M domain values for different operations. It is clear that the DB owner processing time is significantly less than the server processing time. In case of 5M (20M) OK values and 50 DB owners, each DB owner took at most 4s (13s) in PSI Sum (PSI Sum) query, while servers took at least 20s (72s) in PSI (PSI) query; see Figure~\ref{fig:Prism dealing with multiple DB owners}.

\smallskip
\noindent
\noindent\textbf{Exp 4. Impact of bucketization.}
Figure~\ref{fig:bucketization} shows the reduction in the number of values on which we need to execute PSI when using  bucketization technique (\S\ref{subsec:Bucketization-based PSI}).
We created a tree with fanout 10, height 9, and 100M values at the leaf level. In Figure~\ref{fig:bucketization}, we refer to the percentage of leaf nodes of the tree that containing one as \emph{fill factor}. We use a term \emph{actual domain size}  (in Figure~\ref{fig:bucketization})  that refers to the number of items on which we execute PSI. The actual domain size is different from the \emph{real domain size} that refers to domain values given to us, \textit{i}.\textit{e}., 100M. The actual domain size depends on the fill factor and impacts the performance of PSI. When fill factor is 100\% (\textit{i}.\textit{e}., all leaf nodes have one; thus, the entire tree has one), the actual domain size was 111M. But, if the fill factor was only 0.01\% of 100M values (\textit{i}.\textit{e}., 10K), then most of the tree contained zero; thus, we run PSI only on actual domain size of 400K, instead of real domain size of 100M. Note, for this experiment, we generated the data randomly. If there is a correlation in the data (the case in most real-world datasets), bucketization results will be even better.

\subsection{Comparing with Other Works}
\label{subsec:Comparing with Other Works}
\BB
We compare \textsc{Prism} against the state-of-the-art cloud-based industrial MPC-based systems: Galois Inc.'s Jana~\cite{DBLP:journals/iacr/ArcherBLKNPSW18}, since it provides identical security guarantees at servers as \textsc{Prism}. To evaluate Jana, we inserted two LineItem tables (each of 1M rows) having $\langle$OK, PK, LN, SK, DT$\rangle$ columns and executed join on OK column. However, the join execution took more than 1 hour to complete.

\cite{DBLP:conf/ccs/Kerschbaum12,DBLP:conf/ic2e/LiuNZGH14,DBLP:journals/tcc/QiuLSLW18,DBLP:journals/tdsc/AbadiTMD19,DBLP:conf/fc/AbadiTD16,DBLP:conf/fc/KamaraM0S14,DBLP:conf/sac/Kerschbaum12} provide \textbf{cloud-based PSI/PSU/aggregation} techniques/systems. \textbf{\emph{We could not experimentally compare \textsc{Prism} against such systems}}, since none are open-source, except SMCQL~\cite{SMCQL}, (which we installed and works for a very small data and incurs runtime errors). Thus, in Table~\ref{tab:prev_papers_table}, we report experimental results from those papers, just for intuition purposes. With the exception of~\cite{DBLP:conf/fc/KamaraM0S14}, none of the techniques supports a large-sized dataset, has quadratic/exponential complexity, or uses expensive cryptographic techniques~\cite{DBLP:journals/tcc/QiuLSLW18}. While~\cite{DBLP:conf/fc/KamaraM0S14} scales better, it does not support aggregation and, also, reveals which item is in the intersection set. For a fair comparison, we report \textsc{Prism} results only for two DB owners in Table~\ref{tab:prev_papers_table}, since other papers do not provide experimental results for more than two DB owners. In our experiments (Figure~\ref{fig:5M rows parallelism increasing DB owner}), \textsc{Prism} supports 50 DB owners and takes at most $\approx$41 seconds on 5M values. Also, note that, in case of 1B values and two DB owners, \textsc{Prism} takes $\approx7.3$mins, unlike~\cite{DBLP:conf/fc/KamaraM0S14} that took $\approx$10mins.

Several \textbf{non-cloud-based PSI approaches} also exist and  \emph{\textbf{cannot be directly compared against}} \textsc{Prism}, due to a different model used (in which DB owners communicate amongst themselves and do not outsource data to cloud) and/or different security properties (\textit{e}.\textit{g}.,\cite{DBLP:conf/ccs/ChenLR17,DBLP:conf/eurocrypt/FreedmanNP04,DBLP:conf/ccs/LiWZ19,DBLP:conf/ndss/HuangEK12,DBLP:conf/ccs/KolesnikovMPRT17,many2012fast,DBLP:conf/ccs/DongCW13,DBLP:conf/stoc/GoldreichMW87,DBLP:conf/ccs/ArakiFLNO16,DBLP:conf/uss/Pinkas0Z14}). Many schemes including Yao's approach~\cite{DBLP:conf/focs/Yao86} for comparison/max finding were proposed; \textit{e}.\textit{g}.,~\cite{DBLP:conf/tcc/DamgardFKNT06,DBLP:conf/pkc/NishideO07,sepia,DBLP:conf/icisc/HamadaKICT12,DBLP:conf/nordsec/BogdanovLT14,DBLP:conf/icde/VaidyaC05,DBLP:conf/icccn/BurkhartD10}. Such techniques have limitations: many communication rounds, restricted to two DB owners, quadratic computation cost at servers, not dealing with malicious adversaries in cloud settings, and/or no support for result verification.

\smallskip
\noindent
\textbf{Comparison between \textsc{Prism} and \textsc{Obscure}~\cite{DBLP:journals/pvldb/GuptaLMP0A19}.} While both \textsc{Prism} and \textsc{Obscure} are based on secret-sharing, they are significantly different from each other in terms of:
(\textit{i}) purposes: \textsc{Prism} is for computing simple aggregation over PSI/PSU queries over multi-owner databases, while \textsc{Obscure} is for complex conjunctive/disjunctive aggregation query processing over outsourced data and does not support PSI/PSU queries;
(\textit{ii}) implementation: \textsc{Prism} is based on domain-based representation, while \textsc{Obscure} is based on unary representation;
(\textit{iii}) query execution complexities: \textsc{Prism} complexity is $\mathcal{O}(m\times \mathrm{Dom}(A_c))$, where $m$ is \#DB owners and $\mathrm{Dom}(A_c)$ is the domain of  attribute $A_c$, while \textsc{Obscure} complexity is $\mathcal{O}(n\times L)$, where $n$ is the number of tuples and $L$ is the length of a value in unary representation.
Thus, \textbf{\emph{a direct comparison between the two non-identical systems is infeasible}}. Full version~\cite{full_version} shows overheads of these different secret-sharing techniques.

\section{Conclusion}
\label{sec:Conclusion}
\BB
This paper describes \textsc{Prism} based on secret-sharing that allows multiple DB owners to outsource data to (a majority of) non-colluding servers, behaving like honest-but-curious and malicious servers in terms of computations that they perform. \textsc{Prism} exploits the additive and multiplicative homomorphic property of secret-sharing techniques to implement set operations (intersection, union) and aggregation functions. Experimental results show \textsc{Prism} scales to both a large number of DB owners and to large datasets.



\end{document}